\documentclass[iop,apj]{emulateapj}	
\usepackage{subfigure}
\usepackage{times}
\usepackage{enumerate}
\usepackage{enumitem}

\newcommand{\subfloat}{\subfigure} 

\newcommand{\gray}{$\gamma$-ray}
\newcommand{\gRay}{$\gamma$-Ray}
\newcommand{\rigidity}{\rho}
\newcommand{\Rbr}{\rigidity_\mathrm{br}}
\newcommand{\Rinj}{\rigidity_\mathrm{inj0}}
\newcommand{\Rinjbr}{\rigidity_\mathrm{inj1}}
\newcommand{\Rez}{\rigidity_\mathrm{e0}}
\newcommand{\Reo}{\rigidity_\mathrm{e1}}
\newcommand{\Ret}{\rigidity_\mathrm{e2}}
\newcommand{\gez}{g_\mathrm{e0}}
\newcommand{\geo}{g_\mathrm{e1}}
\newcommand{\get}{g_\mathrm{e2}}
\newcommand{\geh}{g_\mathrm{e3}}
\newcommand{\Rdiflow}{\rigidity_\mathrm{0}}
\newcommand{\Rdifhigh}{\rigidity_\mathrm{1}}
\newcommand{\lbr}{\Lambda_\mathrm{br}}

\newcommand{\scenario}{scenario}
\newcommand{\scenarios}{scenarios}

\newcommand{\scenR}{{\it Scenario~R}}
\newcommand{\hypS}{{\it Hypothesis~S}}
\newcommand{\hypA}{{\it Hypothesis~A}}
\newcommand{\scenP}{{\it Scenario~P}}
\newcommand{\scenIbr}{{\it Scenario~I~${(a)}$}}
\newcommand{\scenIcomp}{{\it Scenario~I~${(b)}$}}
\newcommand{\scenL}{{\it Scenario~L}}
\newcommand{\scenH}{{\it Scenario~H}}

\newcommand{\calc}[1]{{\it Calculation~#1}}
\newcommand{\calcs}[1]{{\it Calculations~#1}}

\newcommand{\calculations}{calculations}
\newcommand{\calcR}{{\it Calculation~R}}
\newcommand{\calcS}{{\it Calculation~S}}
\newcommand{\calcP}{{\it Calculation~P}}
\newcommand{\calcI}{{\it Calculation~I}}
\newcommand{\calcL}{{\it Calculation~L}}
\newcommand{\calcH}{{\it Calculation~H}}

\newcommand{\fermilat}{{\it Fermi}-LAT}
\newcommand{\galprop}{GALPROP}
\newcommand{\valf}{v_{\rm Alf}}
\newcommand{\Dpp}{D_{pp}}
\newcommand{\Dxx}{D_{xx}}
\newcommand{\ddp}{\frac{\partial}{\partial p}}

\newcommand{\hi}{H~{\sc i}}
\newcommand{\hii}{H~{\sc ii}}
 
\newcommand{\be}{\begin{equation}}
\newcommand{\ee}{\end{equation}}
\newcommand{\DeltaBreak}{\Delta_\mathrm{br}}
\newcommand{\DeltapHe}{\Delta_{p/\mathrm{He}}}
\newcommand{\pbar}{$\bar{p}$}
\newcommand{\pbarit}{$\bar{p}$}

\slugcomment{}

\shorttitle{Testing the origin of HECRs}
\shortauthors{Vladimirov~et~al.}

\begin{document}

\title{TESTING THE ORIGIN OF HIGH-ENERGY COSMIC RAYS}
 
\author{
A.~E.~Vladimirov\altaffilmark{1},
G.~J\'ohannesson\altaffilmark{2},
I.~V.~Moskalenko\altaffilmark{1,3}, and
T.~A.~Porter\altaffilmark{1}
}

\altaffiltext{1}{W. W. Hansen Experimental Physics Laboratory
Stanford University, Stanford, CA 94305, USA}
\altaffiltext{2}{Science Institute, University of Iceland, Dunhaga 5, IS-107 Reykjav\'{i}k, Iceland}
\altaffiltext{3}{Kavli Institute for Particle Astrophysics and Cosmology, 
Stanford University, Stanford, CA 94305, USA}

\begin{abstract}
Recent accurate measurements of cosmic-ray (CR) protons and nuclei by 
ATIC-2, CREAM, and PAMELA reveal: 
a) unexpected spectral hardening in the spectra of CR species above a few 
hundred GeV per nucleon, 
b) a harder spectrum of He compared to protons, and 
c) softening of the CR spectra just below the break energy. 
These newly-discovered features may offer a clue to the origin of the 
observed high-energy Galactic CRs. 
We discuss possible interpretations of these spectral features and 
make predictions for the secondary CR fluxes and secondary to primary
ratios, anisotropy of CRs, and 
diffuse Galactic \gray\ emission in different phenomenological scenarios. 
Our predictions can be tested by currently running or near-future high-energy 
astrophysics experiments.

\end{abstract}


\section{Introduction}\label{introduction}

The
spectrum of cosmic rays (CRs) has offered few clues to its origin so far. 
The only features observed are at very high and ultrahigh energies 
\citep[see, e.g., Figure~1 in][]{Swordy2001}: the so-called knee at a 
few times $10^{15}$ eV \citep[][]{knee1958,Haungs2003}, the second 
``knee'' at $\sim$$10^{18}$ eV, the ``ankle'' at higher 
energies \citep[][]{Abbasi2005}, and a spectral steepening above 
$10^{20}$ eV \citep{2009APh....32...53T,AUGER2010}. 
Because of the limited size of Galactic accelerators and strength of magnetic
fields in the acceleration region (e.g., in supernova remnants [SNRs]),
it is believed 
that the CRs below the knee are Galactic, while above the knee they have an 
extragalactic origin, with the knee itself 
being due to propagation effects 
and a transition between the two populations of 
CRs \citep[][]{Berezinskii1990, SMP2007}. 

The power-law spectrum below the knee is thought to be the result of CR 
acceleration in SNR shocks \cite[see, e.g.,][]{Drury2001},
which is steepened  to the observed index $\sim 2.75$
by propagation in the interstellar medium (ISM) and 
eventual leakage from the Galaxy. 
The interstellar diffusion coefficient is typically assumed to be a 
power law in particle rigidity, based on numerous studies of 
magnetohydrodynamical turbulence \cite[see, e.g.,][]{Biskamp2003}.
The turbulent cascade often leads to a distribution of magnetic 
energy that is well described by a power law. 
For energies below $\sim20$~GeV~nucleon$^{-1}$, 
the CR spectrum flattens due to the 
modulation in the heliosphere --- a combined effect of the solar wind and 
heliospheric magnetic field. 
Measurements of CR composition below a few GeV~nucleon$^{-1}$ 
offer detailed information on elemental and isotopic 
abundances \citep[][]{HEAO3, ACE-CRIS-2001, TRACER2011}, including the peaked shape of 
the secondary-to-primary nuclei ratio (e.g., B/C, sub-Fe/Fe) and abundances 
of long-lived radionuclides (such as $^{10}$Be, $^{26}$Al, $^{36}$Cl, 
and $^{54}$Mn). 
These measurements are used to derive the (model-dependent) diffusion 
coefficient and the size of the Galactic volume filled with CRs
\citep{SM1998,Ptuskin1998,Webber1998}, the so-called halo. 
Models of CR propagation are in reasonable agreement with available 
data \citep[e.g.,][]{SMP2007, Trotta2011}, with a few exceptions, 
including the unexpected rise in the positron fraction observed by 
PAMELA \citep[][]{PAMELA2009positrons}.

The data recently collected by three experiments, 
ATIC-2 \citep[][]{ATIC2008elements, Panov2009}, 
CREAM \citep{CREAM2010break, CREAM2011pHe}, and PAMELA \citep[][]{PAMELA2011}, 
indicate a break (hardening) of the spectra of the most abundant CR species 
above a rigidity of a few hundred GV. 
The break rigidity, $\Rbr$, is best measured by PAMELA and occurs 
at approximately the same rigidity for $p$ and He, $\Rbr = 240$~GV. 
The PAMELA data for $10$ GV $\la \rigidity < \Rbr$ agree very well 
with the earlier data from AMS and BESS (see \citealt{AMS2000, BESS2004} 
and Figure~1 of \citealt{PAMELA2011}), while ATIC-2 data points 
for $\rigidity < \Rbr$ are somewhat lower. 
We take the PAMELA data as the most accurate for $\rigidity < \Rbr$. 
For $\rigidity > \Rbr$, ATIC-2 results agree well with those of CREAM.
The change in the spectral index (below/above the break)
is estimated as $\DeltaBreak=\gamma(>\Rbr)-\gamma(<\Rbr)= 0.15$, and 
is the same for protons and He.

Another important feature of the CR spectra discovered by these experiments 
is the difference between the spectral indices of CR protons and He. 
This has been speculated for a long time \citep[e.g.,][and references therein]{Biermann1995}, but the experimental uncertainties were too large to be conclusive 
\citep[see the collection of CR proton and He measurements in][]{Moskalenko2002}. 
The new measurements by the ATIC-2, CREAM, and PAMELA experiments 
confirm this with high significance.
The spectrum of He is found to be harder than the spectrum of protons 
for energies up to, at least, 10$^4$~GeV~nucleon$^{-1}$.
The difference between the proton and He spectral indices calculated 
by \cite{PAMELA2011} using the PAMELA data is $\Delta \gamma = 0.10$, and 
it is approximately the same above and below $\Rbr$.
Within the statistical and systematic uncertainty, the measured $p$/He flux 
ratio appears to be a smooth function of rigidity, continuous at $\Rbr$. 
This shows that the difference in the spectral slope of protons and He nuclei 
persists into the ultra-relativistic regime.

There is also fine structure in the spectra that may provide some 
clues to the nature of the observed features:
PAMELA data clearly show a spectral softening at the break rigidity (which 
we refer to as the ``dip'', below). 
\cite{PAMELA2011} have shown the softening to be statistically 
significant at the 95\% confidence level for the spectra as functions 
of particle rigidity, and at the 99.7\% level for the same data in terms of 
kinetic energy per nucleon. 
The softening is more pronounced in the He spectrum. 

Rather than proposing a detailed interpretation of the observed features, 
in this paper we discuss broad categories of models, 
hereafter called \emph{Scenarios}, and propose 
their observational tests.
A particular realization of each \scenario\ is called \emph{Calculation}.
The quantitative analysis is done using
the GALPROP code\footnote{The project Web site http://galprop.stanford.edu/} 
\citep[][]{SM1998}.

We study the interpretations of the $p$/He ratio variation separately from
the interpretations of the spectral break at $\Rbr$. In Section~\ref{refcase},
we introduce the reference scenario based on the pre-PAMELA data.
In Section~\ref{subsectionphe}, we discuss possible explanations
of the $p$/He ratio decline with energy: inherent nature of CR sources and
spallation effects. Section~\ref{breakdip} presents four physical scenarios
that could lead to the observed spectral break at $\Rbr$: injection effects,
propagation effects, and local low- or high-energy CR source.
The framework of our CR propagation calculations is described in
Sections~\ref{galprop} and \ref{dr_and_pd}, and specific calculation setups in Section~\ref{models}.
The sections following that discuss our results and their implications to
CR observations and CR propagation modeling.

\section{Scenarios}\label{classes}

\subsection{Reference Case}\label{refcase}

{\it \scenR: {\underline R}eference scenario}. First, we introduce
a reference case based on the pre-PAMELA data. 
In this scenario, the CR injection spectrum above 10~GV is a single 
power law up to the ``knee'' in the CR spectrum, with the same spectral 
index for all CR species. 
The rigidity dependence of the diffusion coefficient 
at high energies is also taken as a 
single power law for all energies.
The CR source distribution is described in Section~\ref{calculations}.

\scenR\ provides reasonable agreement with the pre-PAMELA data, but it 
cannot reproduce the spectral features evident in the new data 
discussed in this paper: the 
difference between proton and He spectra, the spectral break, or the dip. 
Below, we describe several broad categories of models that encompass viable 
explanations for these new features.
The comparison of predictions for these other models for quantities other than CR 
proton, He, and electron spectra with predictions of \calcR\ qualitatively 
illustrates the significance of the difference between different \scenarios. 

\subsection{$p$/He Ratio: Acceleration and Spallation Hypotheses}\label{subsectionphe}

The confirmation of a significant difference between proton and He spectral indices 
poses a challenge for theories of CR acceleration and propagation. 
Whatever the physical cause of this difference in spectra may be, it seems 
to affect heavier nuclei in the same way as it does He 
\citep[see, e.g.,][for spectra of nuclei]{CREAM2009elements}, giving them 
a harder spectrum than that of protons.

Diffusive shock acceleration (DSA) predicts the spectrum of He 
{\it trapped} in a shock to be 
harder than that of protons due to its lower $Z/A$ ratio, 
but only for non-relativistic energies \citep[e.g.,][]{EDM1997}.
However, in particles {\it escaping} from a shock, $p$/He ratio 
may decline with energy, if DSA is rapid, and 
the injection of He into the acceleration process varies in a way that 
enhances He acceleration in young shocks. 
This could happen due to the inherent property of particle
injection in shocks
\citep{Malkov2011}, or if the abundance of He \citep{OhiraIoka2011}
or magnetic field orientation \citep{Biermann1995} is
inhomogeneous in the SNR environment.
We encompass these mechanisms into the Acceleration
Hypothesis (\hypA), which is discussed below.
Note that propagation effects may contribute to the $p$/He 
spectral difference because
the second-order Fermi process (reacceleration) in the interstellar 
medium makes the He spectrum harder due to its lower $Z/A$ ratio.
This effect, however, 
does not extend to the ultra-relativistic regime \citep[e.g.,][]{SMP2007}. 

An alternative idea, suggested by \cite{BA2011a}, is that
that spallation of CR nuclei ($Z>1$) may lead to 
hardening of their spectra. 
This is because
the lower energy CRs have longer confinement times in
the Galaxy, 
and their flux is depleted by spallation more than the flux
of higher energy nuclei.
Hardening occurs only
if the spallation timescale is short compared to the
confinement timescale of the nuclei.
Note that \cite{BA2011a} consider CRs at energies above 1~TeV
and do not attempt to make their model
consistent with the CR data at low energies, where various effects,
such as stochastic reacceleration and significant production of
secondaries, come into play.
We investigate this idea, extending it to lower energies, and hereafter refer to it as \hypS.
Our calculations include spallation of all nuclei species at all energies, the default 
with GALPROP. 
However, as our results for \calcR\ show (see Section~\ref{results}), the 
effect of spallation on the He spectrum is insignificant, and the $p$/He ratio 
above 10~GV is flat. 
Below, we demonstrate that, with some model tuning (i.e., \calcS$_{1,2}$), fragmentation 
may indeed lead to hardening of the He spectrum. 
We also assess the consequences of the required model modifications.

Note that in this section we refer to the results of our calculations for
\hypS . The framework and details of these calculations are
formally introduced later in the text in Section~\ref{calculations}. However, we would
like to briefly discuss \hypS\ here, because in the rest of the paper
we discard \hypS\ and adopt the assumption that the $p$/He decline
is caused by the nature of CR accelerators (\hypA).
The reader interested in our reasoning for discarding the spallation effects hypothesis may find
the explanation in this section. And a detailed description of 
our physical model, computational method and data sources can be found in
Sections~\ref{calculations} and \ref{results}, where we discuss
the spectral break at $\Rbr$.

{\it \hypS: {\underline S}pallation effects}. 
The fraction of fragmented CR nuclear species depends on 
their total inelastic cross section and the effective grammage encountered 
by the CR species in the Galaxy. 
Inelastic cross section fits used in \cite{BA2011a}, taken 
from \cite{HKT2007}, are somewhat larger than those used in our 
calculations \citep{Barashenkov93, BPCODE}. 
Besides that, the gas number density
used in calculations by \citet{BA2011a} yields a significantly larger 
grammage than in our standard models.
To segregate these effects, we construct two calculation setups
for \hypS: \calcS$_{1}$ and
\calcS$_{2}$.
Both calculations adopt the cross section fits 
from \cite{HKT2007}. 
\calcS\ uses parameters similar to that of the reference \calcR,
but a slightly smaller diffusion coefficient to match the
value used by \cite{BA2011a}.
In \calcS$_2$, we additionally 
increase the gas number density relative to \calcR\ by a factor of two.
Note that these calculations use the GALPROP code, which was adapted to 
incorporate the 
above-mentioned inelastic cross sections, 
whereas the production of fragments (daughter isotopes) is calculated 
using a standard set of cross sections and remains unchanged.

\begin{figure}[tbh]
\centering{
\subfloat[][]{
\includegraphics[width=0.490\textwidth]{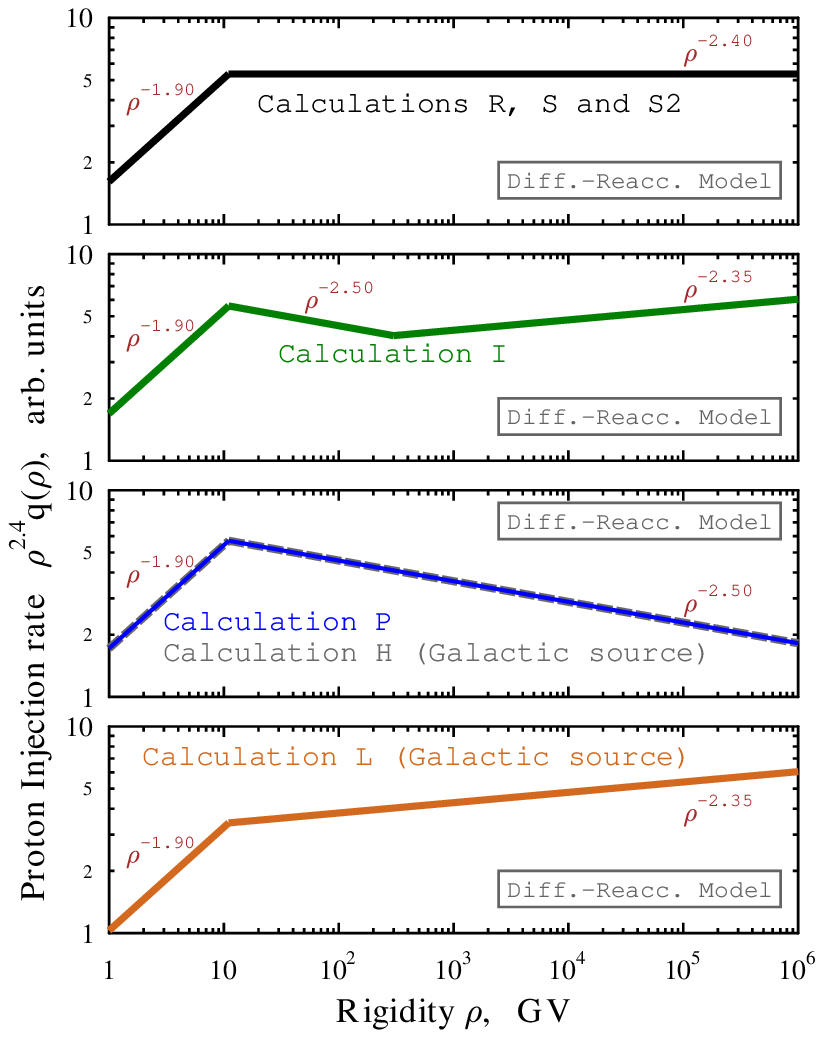}
}
\subfloat[][]{
\includegraphics[width=0.490\textwidth]{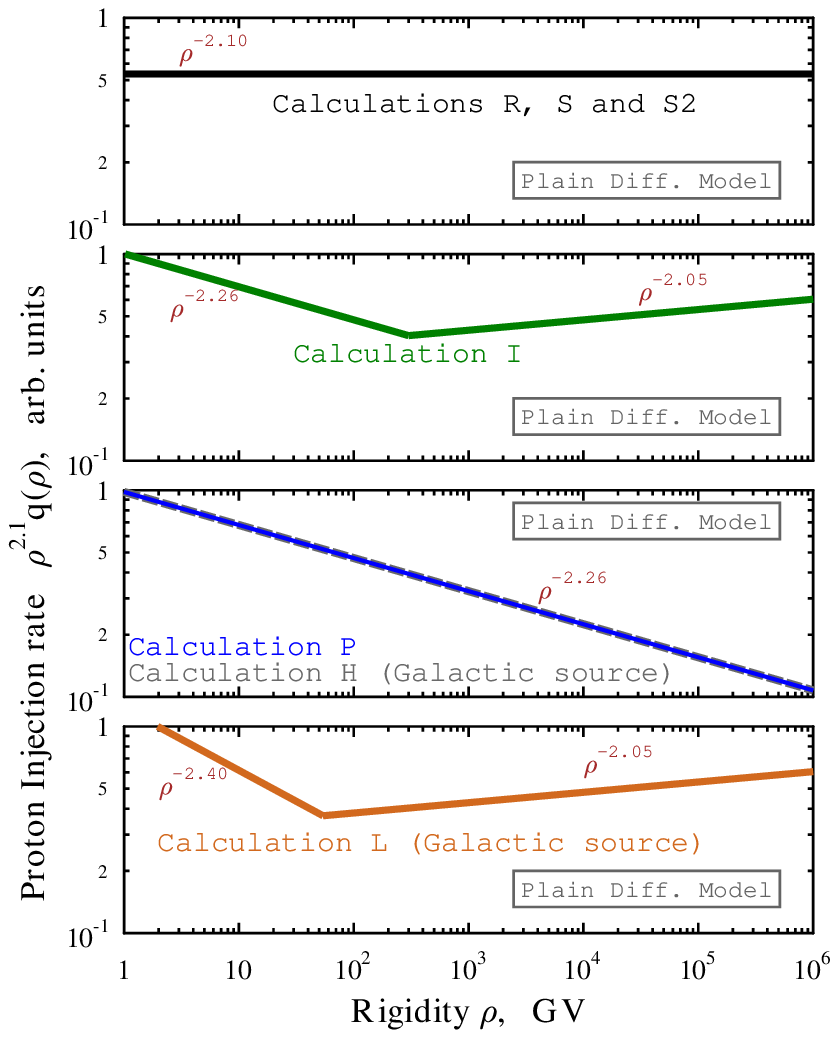}
}
}
\caption{(color in online version) Galactic CR source injection 
spectrum for \calcs{R, S, P, I, L, {\rm and} H} in arbitrary units. 
{\it Left}: diffusive-reacceleration model, {\it right}: plain diffusion model.
The normalization for the injection spectrum was chosen to match local 
measurements of proton and He spectra. 
For all calculations, the lines represent the Galactic CR source injection 
spectrum. ``Local'' sources, present in \calcL\ and \calcH, are not shown here. 
The ``local'' source fluxes at Earth in \calcL\ and \calcH\ were obtained as 
the difference between the observed and propagated Galactic fluxes.}
\label{fig_injection}
\end{figure}

\begin{figure}[tbh]
\centering{
\subfloat[][]{
\includegraphics[width=0.490\textwidth]{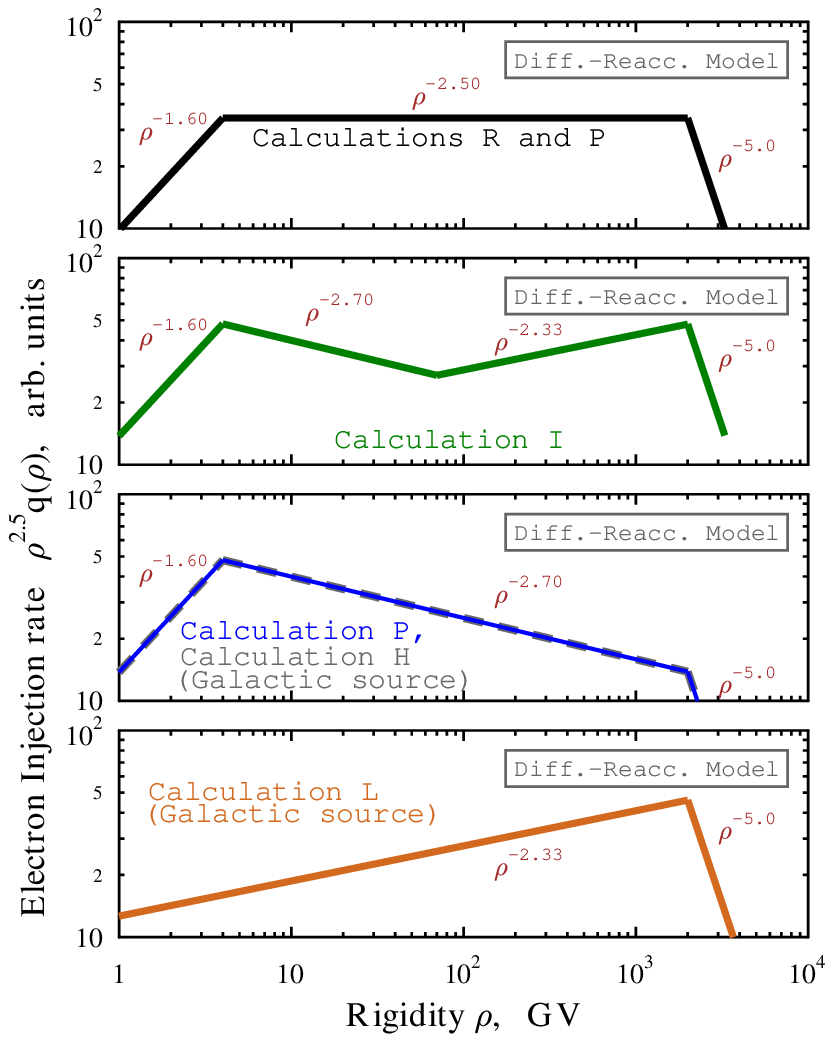}
}
\subfloat[][]{
\includegraphics[width=0.490\textwidth]{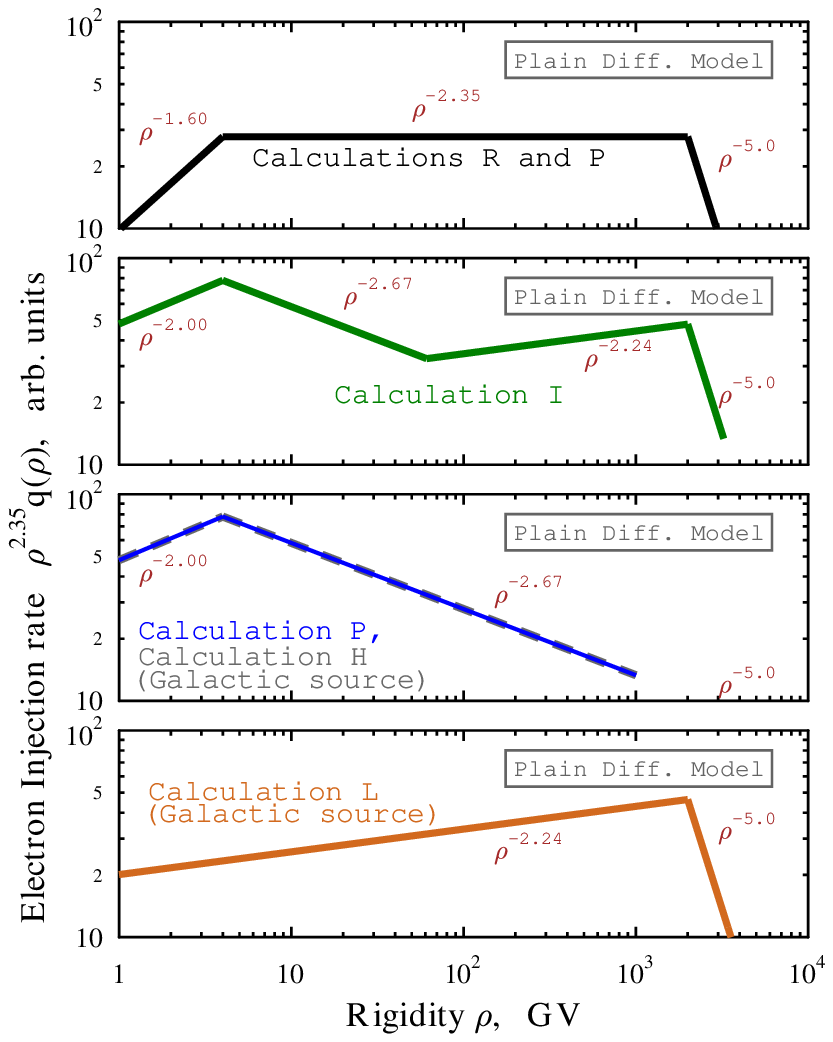}
}
}
\caption{(color in online version) Galactic CR source electron injection 
spectrum for \calcs{R, S, P, I, L, {\rm and} H} in arbitrary units. 
{\it Left}: diffusive-reacceleration model, {\it right}: plain diffusion model.
The normalization for the injection spectrum was chosen to match local 
measurements of proton and He spectra. 
For all calculations, the lines represent the Galactic CR source injection 
spectrum. ``Local'' sources, present in \calcL\ and \calcH, are not shown here. 
The ``local'' source fluxes at Earth in \calcL\ and \calcH\ were obtained as 
the difference between the observed and propagated Galactic fluxes.}
\label{fig_electron_injection}
\end{figure}

\begin{figure}[tbh]
\center{
\subfloat[][]{
\includegraphics[width=0.490\textwidth]{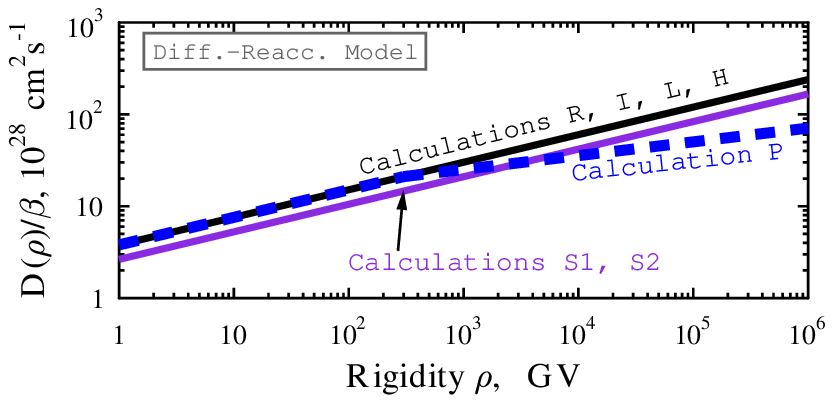}
}
\subfloat[][]{
\includegraphics[width=0.490\textwidth]{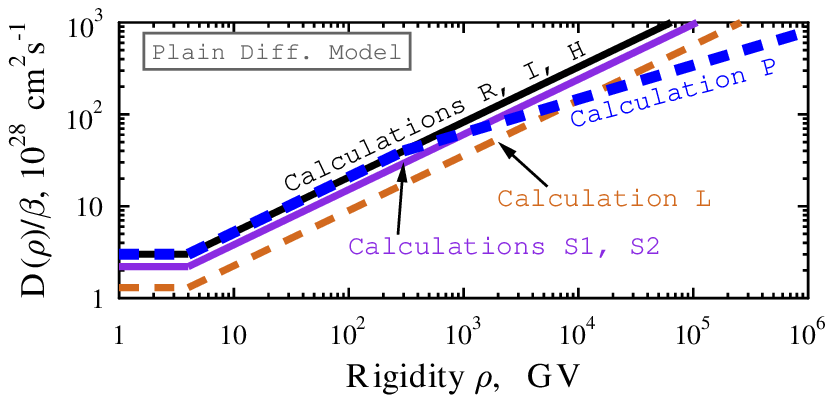}
}
}
\caption{(color in online version) Diffusion coefficient of CRs in the 
Galaxy (see Equation~(\ref{Dxx})). 
{\it Left}: diffusive-reacceleration model, {\it right}: plain diffusion model.
The values of the $D$ coincide for {\it Calculations R, I, L} and {\it H}.
For \calcS, the value of $D$ is slightly smaller at all energies.
For \calcP, a break in the diffusion coefficient is assumed, 
changing $\delta_1/\delta_2=0.30/0.15$ at $\Rdifhigh=300$~GV.}
\label{fig_diffusion}
\end{figure}

Our calculations show that 
with our standard gas distribution based on astronomical 
data \citep{Moskalenko2002}, \calcS$_1$, the amount of hardening is 
insufficient to provide agreement with the PAMELA $p$/He ratio. 
The required hardening of the He ($\mathrm{^3He}+\mathrm{^4He}$) spectrum can be 
achieved only if we assume a 
considerable increase in grammage and simultaneously adopt a set of 
total inelastic cross sections from \cite{HKT2007}, \calcS$_2$.
However, this leads to an overproduction of secondary species in CRs, 
such as antiprotons and boron, so that the calculated B/C ratio does 
not agree with the data. 
The $p$/He ratio obtained in \calcS$_1$ and \calcS$_2$ is
shown in the top panels of Figure~\ref{fig_spallation}
(see Section~\ref{dr_and_pd} for an overview of
the plain diffusion and diffusive-reacceleration models). The calculated B/C
ratio and \pbar\ flux are shown in the middle and bottom panels, respectively.
Read on and see Figures~\ref{fig_injection}--\ref{fig_diffusion} for
details on the parameters of the calculations shown in Figure~\ref{fig_spallation}.

Another important point to consider here is that the measurements 
of PAMELA, ATIC-2, and CREAM are not sensitive to the isotopic composition 
of CR fluxes. 
The He fluxes reported by these experiments and used throughout this 
paper are, in fact, the sum of $^3$He and $^4$He species. 
The dominant channel of $^4$He spallation is the reaction 

\begin{displaymath}
\mathrm{^4He} + \mathrm{p} \rightarrow \mathrm{^3He} + \mathrm{X}.
\end{displaymath}

\noindent
This reaction leads to a hardening of 
the interstellar $^4$He spectrum because lower energy nuclei experience 
more spallation events. 
However, due to production of $^3$He in the same reaction, the {\it total} 
He spectrum does not harden as much as $^4$He alone. 
Further spallation of secondary $^3$He, as well as fragmentation of $^4$He 
into products other than $^3$He, eventually leads to the total He spectrum 
hardening. 
Still, the effect of spallation on the total He spectrum is not as strong 
as on $^4$He alone. 
Equation~(2.1) in \cite{BA2011a} indicates that $^3$He was not included in 
their calculations. 
Therefore, their results would be relevant only for the $p$/$^4$He ratio. 

This is illustrated in Figure~\ref{fig_spallation}, 
where we also plot the ratio of $p$/$^4$He in \calcS$_2$ for reference.
It can be seen that
the overall shape of the $p$/$^4$He ratio matches the measured $p$/He ratio well;
however, a significant fraction of secondary $^3$He changes the shape so that
the calculated $p$/He ratio cannot be adjusted to match the data simultaneously at all rigidities (1 GV -- 10 TV).

{\it \hypA: {\underline A}cceleration effects}.
We have concluded that the adjustments of propagation model required to 
reproduce the observed $p$/He ratio 
in \hypS$_2$ conflict with the measurements of secondary CR species. 
Therefore, in the rest of this work we adopt an alternative to \hypS,
which we call the ``Acceleration hypotheses''. \hypA\
represents the idea that the nature of CR accelerators
is responsible for a harder spectrum of He than $p$ (see references
at the beginning of this section). 
To incorporate \hypA\ into our calculations, we use an ad hoc
modification for the CR injection spectra.
That is, our calculations assume that nuclei heavier than H are 
injected into the ISM with a harder spectrum than protons. 
The difference between the spectral indices of protons and heavier 
nuclei is the same for all rigidities and is $\DeltapHe=0.07$. 

We do not present a separate calculation for the \hypA\ in this paper.
Instead, we incorporate \hypA\ into calculations that 
study scenarios {\it P, I, L} and {\it H} 
for the break in the $p$ and He spectra (those scenarios are introduced in
Section~\ref{breakdip}).
Figure~\ref{fig_p_he_ratio} shows the $p$/He ratio resulting from this 
modification (see below for parameters of other calculations shown in this figure). 
The figure illustrate that the data reported by PAMELA, 
ATIC-2, and CREAM can be reproduced by including $\DeltapHe=0.07$.

\subsection{Spectral Break and Dip: Propagation, Injection and Local Source Scenarios}\label{breakdip}

We consider the following \scenarios\
for an explanation of the break at $\Rbr$ and the dip just below $\Rbr$:
(1) interstellar propagation effects, 
(2) modification of CR injection spectrum at the sources, 
(3) composite Galactic CR spectrum,
(4) effects of local sources at low energies ($\rho < \Rbr$), and
(5) effects of local sources at high energies ($\rho > \Rbr$). 
Particular realizations of these \scenarios\ (\calculations) are discussed in detail in Section~\ref{models}; 
their parameters are summarized in Tables~\ref{table1} and \ref{table1PD}.

{\it \scenP: interstellar {\underline P}ropagation effects}. 
Transport of CRs in the ISM is subject to considerable 
uncertainties, because the properties of interstellar magnetic turbulence 
are not very well known \citep[][]{ElmegreenScalo2004, ScaloElmegreen2004}. 
This makes CR observations a valuable indirect probe of quantitative 
features of particle transport (e.g., the diffusion coefficient, $D$) 
in the Galaxy.
Therefore, in this scenario, the break in the observed proton and He 
spectra is attributed to a change in CR transport properties at 
rigidity $\Rbr$. 
This scenario is represented by \calcP, which has a break in the rigidity 
dependence of the diffusion coefficient at $\rigidity = \Rbr$. 
For $\rigidity < \Rbr$, we use 
the functional form of $D(\rigidity)$ obtained in the earlier 
comprehensive analysis of CR data by \citet{Trotta2011}, and 
for $\rigidity > \Rbr$, we adjust the rigidity dependence of $D(\rigidity)$ 
to match the observations of PAMELA, ATIC-2, and CREAM, as discussed above.

{\it \scenIbr: CR {\underline I}njection effects, source with a spectral break interpretation}.
Existing models of CR production by SNR 
shocks \cite[e.g.,][]{CAB2010, PZS2010} predict a smooth spectrum of 
CR particles injected by an SNR into the Galaxy. 
Such models usually consider a shock in a semi-infinite medium or 
assume spherical symmetry.
The spectrum predicted by these models may gradually harden with energy 
between 10~GeV and 100~TeV, but not as rapidly as in the PAMELA data.
Note that particle transport, magnetic turbulence generation, and nonlinear 
feedback of particles and magnetic fields on shock structure are not 
strictly constrained in these models. 
The spectrum of particles leaking from an SNR shock has never been 
observed directly. 
It is therefore conceivable that with some parameter tuning, present models 
of particle acceleration may predict a more pronounced hardening in the 
spectrum of particles injected into the ISM, consistent 
with the new data. 
Alternatively, particle acceleration models that take into account 
the asymmetry of SNRs may predict a break in the particle spectrum produced 
by a single SNR. 
For example, in the model of \cite{Biermann2010}, the break, or upturn, 
occurs due to the contribution of the SNR's polar cap.
This case, hereafter referred to as \scenIbr, is represented by \calcI, 
which features a Galaxy-wide source spectrum with a hardening at $\Rbr$.
The diffusion coefficient does not have a break in this scenario.

{\it \scenIcomp: CR {\underline I}njection effects, composite source interpretation}. While SNRs (isolated or in superbubble regions) are believed to be 
the primary sources of Galactic CRs, different classes of supernovae and 
their environments, as well as other CR sources, can combine to produce 
the observed CR spectrum. 
Generally speaking, different types of CR sources could have different 
spatial distributions throughout the Galaxy.
In this work, we make the simplifying assumption (1) that there are only two types of 
CR sources, and (2) the spatial distributions are the same for both types 
of CR sources. 
If one source dominates the low energy part of the CR spectrum, and 
the other the high energy part, then \calcI, with a hardening of the 
Galactic CR source at $\Rbr$, adequately encompasses this composite 
source scenario as well. 
This scenario may be generalized to a distribution
of CR sources with different parameters. \cite{Yuan2011} have shown 
that in general, dispersion in the CR source spectral indices
results in the concavity of the observed CR spectrum.
We use the same computational setup to calculate the observed quantities 
for \scenIbr\ and \scenIcomp, and we call it just \calcI.
A subtle advantage of the composite source interpretation 
of \calcI\ (i.e., in \scenIcomp) is its ability to explain the dip more 
naturally than the source with an inherent break 
scenario (see the discussion of the dip in Section~\ref{results}).

{\it \scenL: local {\underline L}ow-energy source}. 
This scenario encompasses interpretations that assume that the observed 
spectral break is caused by a local source dominating the CR spectrum 
at {\it low} rigidities, $\rigidity < \Rbr$. 
Unlike \scenIcomp, the present scenario assumes that the low-energy source 
is not typical for the Galaxy as a whole. 
This scenario is formulated as \calcL, in which the Galactic CR spectrum 
is hard, matching the observations 
of PAMELA, ATIC-2, and CREAM for $\rho>\Rbr$. 
For $\rho<\Rbr$, the flux of Galactic CRs is lower than the observed flux, 
and we assume that the difference is accounted for by the hypothetical local 
source. 
We assume the extreme case of a very local low energy source. 
This means that we do not calculate propagation of CRs from that source
and only the Galactic sources with the hard spectrum are used to calculate
the production of secondaries and the diffuse Galactic \gray\ emission.
This scenario contrasts with \scenIcomp, where
the sources of low-energy CRs are distributed across the Galaxy.
The case of intermediate local source extent falls in between \scenL\ 
and \scenIcomp.

{\it \scenH: local {\underline H}igh-energy source}. This scenario is 
analogous to \scenL, but with Galactic sources dominating the CR flux 
for $\rigidity < \Rbr$, and the spectral break produced by a 
local {\it high-energy} source dominating the observed flux 
for $\rigidity>\Rbr$. 
The calculation representing this scenario is referred to as \calcH. 
The assumption of the high-energy source being very local is made in this 
calculation identically to how it was done in \calcL, i.e., the production
of secondaries and the diffuse 
Galactic \gray\ emission is determined solely by the Galactic CR sources.

\section{Calculations}\label{calculations}

\subsection{GALPROP Code}\label{galprop}

The \galprop{} project began in late 1990s \citep{SM1998} 
and has been in continuous development since. 
The code is available 
from the dedicated Web site where a facility for users to run the 
code via online forms in a web 
browser\footnote{http://galprop.stanford.edu/webrun/} is also 
provided \citep{GalpropWebrun}.

The \galprop{} code solves the CR transport equation for a given source 
distribution and boundary conditions for all CR species. 
This equation includes diffusion, 
a galactic wind (convection), diffusive reacceleration in the
ISM, energy losses, nuclear fragmentation, 
radioactive decay, and production of secondary particles and isotopes:

\begin{eqnarray}
\label{eq.1}
\frac{\partial \psi}{\partial t} 
&=& q({\mathbf r}, p)+ \nabla \cdot ( \Dxx\nabla\psi - {\mathbf V}\psi )
+ \ddp\, p^2 \Dpp \ddp\, \frac{1}{p^2}\, \psi \nonumber\\
&-& \frac{\partial}{\partial p} \left[\dot{p} \psi
- \frac{p}{3} \, (\nabla \cdot {\mathbf V} )\psi\right]
- \frac{1}{\tau_f}\psi - \frac{1}{\tau_r}\psi\, ,
\end{eqnarray}

\noindent
where $\psi=\psi ({\mathbf r},p,t)$ is the CR number density per unit total
particle momentum (i.e., $\psi(p)dp = 4\pi p^2 f(p) dp$ in terms of
phase-space density $f(p)$), $q({\mathbf r}, p)$ is the source term,
$\Dxx$ is the spatial diffusion coefficient, ${\mathbf V}$ is the
convection velocity, reacceleration is described as diffusion in
momentum space with diffusion coefficient $\Dpp$,
$\dot{p}\equiv dp/dt$ is the momentum loss rate, $\tau_f$ is the time
scale for fragmentation, and $\tau_r$ is the timescale for
radioactive decay. 
The numerical solution of the transport equation is
based on a Crank-Nicholson \citep{Press1992} implicit second-order
scheme. 
The spatial boundary conditions assume free particle escape, e.g.,  
$\psi(R_h,z,p) = \psi(R,\pm z_h,p) = 0$,
where $R_h$ and $z_h$ are the boundaries for a cylindrically symmetric geometry.

The source function $q$ is

\begin{equation}
\label{eq.3}
q({\mathbf r}, \rigidity) = q_\mathrm{pri} ({\mathbf r}, \rigidity) + \sum q_\mathrm{sec}({\mathbf r}, \rigidity),
\end{equation}

\noindent
where $q_\mathrm{pri}$ represents the primary CR sources, 
and the $q_{\mathrm{sec}}$ term is for the 
sources of secondary isotopes (i.e., nuclear reactions in the ISM), and
$\rho \equiv pc/Ze$ is the magnetic 
rigidity where $p$ is momentum and $Ze$ is the charge.
The distribution of primary Galactic CR sources used in this work is based 
on the supernova distribution from \cite{CB1998}.

While GALPROP's numerical scheme can accommodate arbitrary energy dependence
of CR source function, in this work we parameterize the source function of primary nuclei
as a broken power law in particle rigidity:

\begin{eqnarray}
&&q_\mathrm{pri} ({\mathbf r}, \rigidity) \propto Q({\mathbf r}) \left( \frac{\rigidity}{\Rinj} \right)^{g},\\
&&g=\left\{
\begin{tabular}{l}
$g_0\ {\rm for}\ \rho < \Rinj$ \nonumber\\
$g_1\ {\rm for}\ \Rinj \le \rho < \Rinjbr$ , \nonumber\\
$g_2\ {\rm for}\  \rho \ge \Rinjbr$
\end{tabular} \right. \nonumber
\end{eqnarray}

\noindent
The source function for primary CR leptons is similar to that of nuclei, but with up to two breaks.

Likewise, the spatial diffusion coefficient is given by
\begin{eqnarray}
D_{xx} &=& \beta D_{0}  \left( \frac{\rigidity}{\Rdiflow}\right)^\delta, \\
\delta&=&\left\{
\begin{tabular}{l}
$\delta_0\ {\rm for}\ \rho < \Rdiflow$ \nonumber\\
$\delta_1\ {\rm for}\ \Rdiflow \le \rho < \Rdifhigh$ , \nonumber\\
$\delta_2\ {\rm for}\  \rho \ge \Rdifhigh$
\end{tabular} \right. \nonumber
\label{Dxx}
\end{eqnarray}

\noindent
where $D_{0}$ is the normalization at rigidity $\Rdiflow$ and $\beta \equiv v/c$. 
The power-law index $\delta=1/3$ corresponds to Kolmogorov diffusion (see 
Section~\ref{galprop}). 

GALPROP solves the time-dependent propagation equation, and
in this work, the steady-state solutions of Equation~(\ref{eq.1}) are
obtained assuming that the source functions are time-independent
and integrating the equation over a long enough time interval. 
The accelerated solution technique is used, where the initial 
time step, $\Delta t=10^9$~yr, is large compared to the propagation 
timescale, and after $N_s=20$ iterations, $\Delta t$ is reduced by a 
factor of two, etc., until $\Delta t$ becomes small compared to the shortest 
timescale in the system (in our case, 10~yr, to accommodate the 
rapid energy losses of leptons).

The details of physical processes and data used in the \galprop{} code, as 
well as the numerical scheme, can be found elsewhere.
A complete list of relevant publications is available in \cite{GalpropWebrun}; 
the aforementioned \galprop{} Web site contains additional information and 
publications. 

\subsection{Diffusive-Reacceleration and Plain Diffusion Models}\label{dr_and_pd}

Previous studies have shown that the available CR data can be explained in one of the 
two common propagation models:
the {\it diffusive-reacceleration} (D-R) model and the {\it plain diffusion} (PD) model.
These models have  been used in a 
number of studies utilizing the GALPROP code \citep[e.g.,][and references 
therein]{Moskalenko2002,SMR2004,Ptuskin2006,Abdo2009midlatitudes}. 

The D-R model assumes that low energy CRs in the ISM participate in
the second-order Fermi acceleration process. This process is believed
to be caused by stochastic collisions of CR particles with moving
magnetic structures. Averaged in time, such collisions
result in particle diffusion in momentum space with a 
diffusion coefficient $D_{pp}$, which increases
the mean energy of low energy particles. 
If reacceleration is included, $D_{pp}$ is related to $\Dxx$ 
\citep{Berezinskii1990,Seo1994}:

\begin{equation}
\label{eq.2}
\Dpp\Dxx = \frac{4 p^2 \valf ^2}{3\delta(4-\delta^2)(4-\delta) w},
\end{equation}

\noindent
where $w$ characterizes the level of turbulence (we take $w = 1$ because 
only the quantity $\valf ^2 /w$ is relevant),
and $\delta=1/3$ for a Kolmogorov spectrum
of interstellar turbulence
\citep{1941DoSSR..30..301K} or $\delta=1/2$ for a Kraichnan 
cascade \citep{1964SvA.....7..566I, 1965PhFl....8.1385K}, but can also be arbitrary. 
Matching the B/C ratio below 1~GeV in D-R models is known to 
require large values of $\valf$. 
In order to avoid a large bump
in the proton spectrum at low energies, the D-R model requires a break
in the CR injection function around $\rho=10$~GV.

The PD model assumes no reacceleration process,
which corresponds to $\valf=0$. No
break in the CR injection function is required to fit the proton and helium
spectra at low energies in the PD model.
However, in order to fit the B/C data below 1~GeV~nucleon$^{-1}$, the PD model requires a low-energy
break in the diffusion coefficient. Specifically,
 the diffusion coefficient in the PD model must {\it decrease with increasing energy}
below 4~GV in order to fit the B/C measurements below 1~GeV~nucleon$^{-1}$. 
A possible physical justification of such behavior of $D(\rho)$ 
is given by \cite{PMJSZ06} and involves turbulence dissipation
in the ISM.

Each of our calculations (\calcR, {\it P}, {\it I}, {\it L}, and {\it H}) is presented
in two versions: one for the D-R and another for the PD model.

\subsection{Calculation Setups}\label{models}

The parameters of our calculations are summarized in Tables~\ref{table1}
and \ref{table1PD}. 
Figures~\ref{fig_injection}--\ref{fig_diffusion} show the diffusion 
coefficients and the injection spectra used for the different \scenarios. 

\calcR\ is the reference case for this study. The list below outlines the key 
parameters of this calculation for the D-R and PD models.
%
\begin{enumerate}
\item{
For \calcR\ in the D-R model, we chose $g_0=-1.9$, $g_1=g_2=-2.4$ 
(i.e., no break at $\Rinjbr$) 
and $\Rinj=11$~GV for all nucleons, which is consistent with the findings 
of \cite{Trotta2011} based on
a Bayesian analysis of a D-R model using the GALPROP 
code. For the PD model, a low
energy break in the injection spectrum is unnecessary, 
and our \calcR\ uses $g_0=g_1=g_2=-2.1$.
The electron injection spectrum for \calcR\ in the D-R model is similar to that 
from \cite{FermiElectrons2010}, with spectral index in 
rigidity $\Gamma$=1.60/2.50 below/above a break rigidity of $\Rez$=4~GV, and a 
second steepening to $\Gamma$=5.0 above $\Ret$=2~TV.
For the PD model, we use $\Gamma$=2.35 above 4~GeV. Note that
the low-energy break in the electron injection function can be explained
without the corresponding break in the proton injection spectrum, because electrons
and protons may be accelerated in sources via different mechanisms.
}

\item{
For the diffusion coefficient, \calcR\ uses $\delta=0.30$ in the D-R model,
and $\delta=0.60$ in the PD model. The normalization for the diffusion 
coefficient in the D-R model is $D_0=5.75\times 10^{28}$~cm~s$^{-1}$ at $\rigidity=4$~GV, 
which is consistent with the best-fit values obtained by \cite{Trotta2011}. 
For the PD model, we use $D_0=3.0\times 10^{28}$~cm~s$^{-1}$,
in order to match the B/C observations of HEAO, TRACER, and CREAM (see 
Section~\ref{results.bc}). 
In this work, we construct
the PD model with a constant, rather than decreasing,
diffusion coefficient below 4~GV, even though the
model does not reproduce the low energy B/C data. This is done in order to illustrate
the possibility that a local low-energy CR source (\scenL) can simultaneously explain
the break in $p$, He spectra {\it and} fit the low-energy B/C data without requiring 
the diffusion coefficient to decrease with increasing energy (see below).
}
\item{
Finally, to model reacceleration in the D-R model,
we chose $\valf=32$~km~s$^{-1}$, the halo size $z_h=4$~kpc, and the 
normalization of the propagated CR proton spectrum was tuned to the 
observed flux $N_p=10.7\times 10^{-12}$~cm$^{-2}$ s$^{-1}$ sr$^{-1}$ MeV$^{-1}$ 
for $\rigidity = 10^3$~GV. 
These values were obtained by slightly adjusting the best-fit values obtained 
by \cite{Trotta2011} to achieve a good agreement with the PAMELA proton 
spectrum for $\rigidity < \Rinj$. 
These adjusted values are still within one mean square deviation of the 
posterior mean found by \cite{Trotta2011}. 
In all figures, black lines represent the input and output quantities 
pertaining to \calcR.
}
\end{enumerate}

\calcP\ has the same parameters as the reference \calcR, except:
(1) the injection spectrum of protons (electrons) above 
the low energy break $\Rinj$ ($\Rez$) has a softer power-law 
index to give agreement with the PAMELA data below $\Rbr$;
(2) the injection spectra of He and heavier elements ($A>1$) have a power-law 
index harder than that of protons by $\DeltapHe$ for all rigidities
(this represents \hypA);
(3) the rigidity dependence of the diffusion coefficient, 
Equation~(\ref{Dxx}), has a break at rigidity 
$\Rdifhigh$ (i.e., $\delta_1 \neq \delta_2$).
The break in the rigidity dependence of the diffusion coefficient is 
introduced to match the observed break in the CR spectrum.
Besides, we choose $\Rdifhigh \approx \Rbr$ so that $\Rdifhigh$ is slightly 
larger than $\Rbr$ for better agreement with the data for $\rigidity > \Rbr$.
The normalization of the proton flux has been adjusted in \calcP, along with 
the abundance of He, to agree with PAMELA data at all energies (the abundances 
of heavier nuclei were changed by the same factor as He). 
The results for \calcP\ are shown with blue lines in all figures.

\calcI\ differs from \calcR\ in the following ways:
(1) the index of the proton injection spectrum is softer than 
in \calcR\ for $\Rinj<\rigidity<\Rinjbr$;
(2) the injection spectrum has two breaks, i.e., $g_1 \neq g_2$
(this represents \scenIbr\ and \scenIcomp);
(3) electrons also have a softer spectrum for $\Rez<\rigidity<\Reo$,
and a break at $\Reo$,
and (iv) nuclei are injected with a harder spectrum than protons (\hypA). 
This calculation produces a CR spectrum at Earth with a break at $\Rbr$ 
closely matching that of \calcP, but due to a different physical assumption. 
Namely, it is the spectral break in the CR injection spectrum that produces 
the break in \calcI, whereas in \calcP, it occurs because of a break in the 
diffusion coefficient. 
Note that the high-energy break in the electron injection spectrum 
must be stronger than for protons, and it must occur at a lower rigidity than
in protons, in order
to obtain agreement with the electron spectra observed by the \fermilat\ and PAMELA.
The results of this calculation are shown as green lines.

\calcH\ combines two components. 
One component is produced by the Galactic CR sources and propagated 
using GALPROP with the same parameters as \calcI.
This component does not have a break in the CR injection spectrum 
at $\Rinjbr$ and its normalization was tuned
to match the proton and He spectra for $\rigidity < \Rinjbr$. 
Another component is produced by a hypothetical local source, which 
contributes to the total flux only for $\rigidity > \Rbr$. 
We do not calculate CR propagation for the local source; instead, we 
calculate its spectrum at Earth by subtracting the Galactic source spectrum 
from the data of ATIC-2 and CREAM for $\rho > \Rbr$.
As discussed in Section~\ref{results}, this represents the assumption 
that the local source is nearby and not very powerful. 
To compute secondary particles and isotopic ratios in this calculation, we 
assume that isotopic abundances in the local and Galactic sources are 
similar, and that the local source supplies no secondary particles at Earth. 
This lowers, for instance, the B/C and \pbar /$p$ ratios for $\rho > \Rbr$. 
The diffuse \gray\ emission from the Galaxy is calculated 
using only CR fluxes from the Galactic source. 
Gray lines represent this calculation in plots.
 
\calcL\ features a local source that contributes to the {\it low}-energy part
of the CR spectrum ($\rigidity < \Rbr$). The local source is included in the same
way as in \calcH, i.e., its propagated spectrum is calculated as the difference between the
observed CR spectrum and the propagated Galactic component. As in \calcH, 
we do not calculate the propagation of CRs from the local source and assume
that its flux contains no secondary species. 
However, this calculation is very different from all other, because
the {\it low-energy} CRs are a mix of particles which have undergone 
Galactic propagation and recently accelerated particles from the local source. 
Because of that the propagation parameters for this calculation should be estimated 
simultaneously with the parameters of the local source, as in \cite{MSMO03}.  

Indeed, assuming that CRs from the local source are produced so recently that they 
contain no secondary nuclei, and that there is no primary boron in CRs,
one can calculate the B/C ratio for \scenL\ 
in the energy range where the local source contribution is non-negligible.
Matching the B/C ratio in the D-R is quite 
challenging because the local source contribution reduces
the B/C ratio in the range 1--10~GeV~nucleon$^{-1}$.
This reduction may be compensated by assuming
a lower diffusion coefficient in this energy range.
At the same time, the diffusion coefficient
above 10~GeV~nucleon$^{-1}$ cannot be changed very much in order to
maintain agreement with high-energy B/C data.
The above considerations necessitate a larger value of $\delta_1$. 
With a greater $\delta_1$, the diffusive
reacceleration of low-energy protons becomes too strong,
resulting in an overprediction of the proton flux around 1~GeV, and thus $\valf$ should be reduced. 
In fact, we found that in this scenario, the best agreement with 
the PAMELA data for protons is achieved with $\valf=0$ 
(i.e., no reacceleration) and $\delta_0$=0 below a 4~GV (i.e., the PD model). 
This is because any finite $\valf$ hardens the proton spectrum below 1~GeV 
too much to match the PAMELA data. 

Considering the above, we chose to keep \calcL\ in the D-R model
with unchanged diffusion coefficient, in order to illustrate the problem in the B/C fitting.
And in the PD model, the reduction of the B/C ratio below 1~GeV was beneficial for
agreement with data, because in all other PD calculations, B/C
was overpredicted. Quantitatively matching the {\it ACE} data in the PD model
requires the flux of the local source to be relatively small. 
This dictates our choice of a concave Galactic source 
spectrum for the \calcL\ in the PD model, i.e., $g_0<g_1$ (see Table~\ref{table1PD}).
Such a spectrum is similar to the theoretical 
predictions by \cite{PZS2010}. 

Finally, in \calc{S$_1$} (discussed in Section~\ref{subsectionphe})
all parameters are the same as in \calcR, except for the 
diffusion coefficient $D_0$ and the Alfv\'{e}n speed $\valf$. 
$D_0$ is reduced by approximately 25\%, which makes the ratio $z_H/D_0$ equal 
to that in the calculations of \cite{BA2011a}. 
$\valf$ is reduced accordingly to avoid a large bump in the proton spectrum 
at low energies. 
In addition, the propagation calculations use a different set of total 
inelastic cross sections \citep[Equations {[6]--[8]} in][]{HKT2007}. 
\calc{S$_2$} has the same parameters as \calc{S$_1$}, but the density of all 
gas components in the Galactic disk (i.e., \hi, \hii, and H$_2$) is 
multiplied by 2. 
The results of \calc{S$_{1,2}$} are compared with \calcR\ in 
Figure~\ref{fig_spallation}.

\section{Results}\label{results}

The results of {\it Calculations R, S, P, I, L}, and {\it H}, as specified 
in Tables~\ref{table1} and \ref{table1PD}, are summarized in Figures~\ref{fig_cr_spectra} 
through \ref{fig_gamma_ray_midlat}.
Figures~\ref{fig_cr_spectra} and \ref{fig_p_he_ratio} show the proton and 
He spectra and their ratio, and Figure~\ref{fig_electrons} 
shows the CR electron spectrum. 
Figure~\ref{fig_spallation}, illustrating \hypS, is discussed in detail 
in Section~\ref{subsectionphe} and in the figure caption. 

Since the origin of the difference between the slopes of the 
proton and He spectra was discussed in detail in Section~\ref{subsectionphe},
we do not mention that topic in this section, instead concentrating 
on scenarios explaining the spectral break.
{\it Calculations P, I, L} and {\it H} were designed to reproduce the observed 
proton, He and electron spectra and, therefore, cannot
be used to constrain any of these scenarios.
However, their predictions for CR anisotropy and the production of 
secondary species 
(B/C ratio, \pbar\ flux, \pbar/$p$ ratio, $e^{+}$ flux and $e^{+}/(e^{+}+e^{-})$ ratio) differ.
These predictions are shown in Figures~\ref{fig_anisotropy}--\ref{fig_positron_fraction}. 
Predictions for the diffuse Galactic \gray\ emission at intermediate 
latitudes ($10^{\circ}<|b|<20^{\circ}$) are compared with the data 
collected by the \fermilat\ in Figure~\ref{fig_gamma_ray_midlat}. 

\subsection{Proton and He Spectra}\label{results.phe}

\begin{figure*}[tbh]
\center{
\subfloat[][]{
\includegraphics[width=0.490\textwidth]{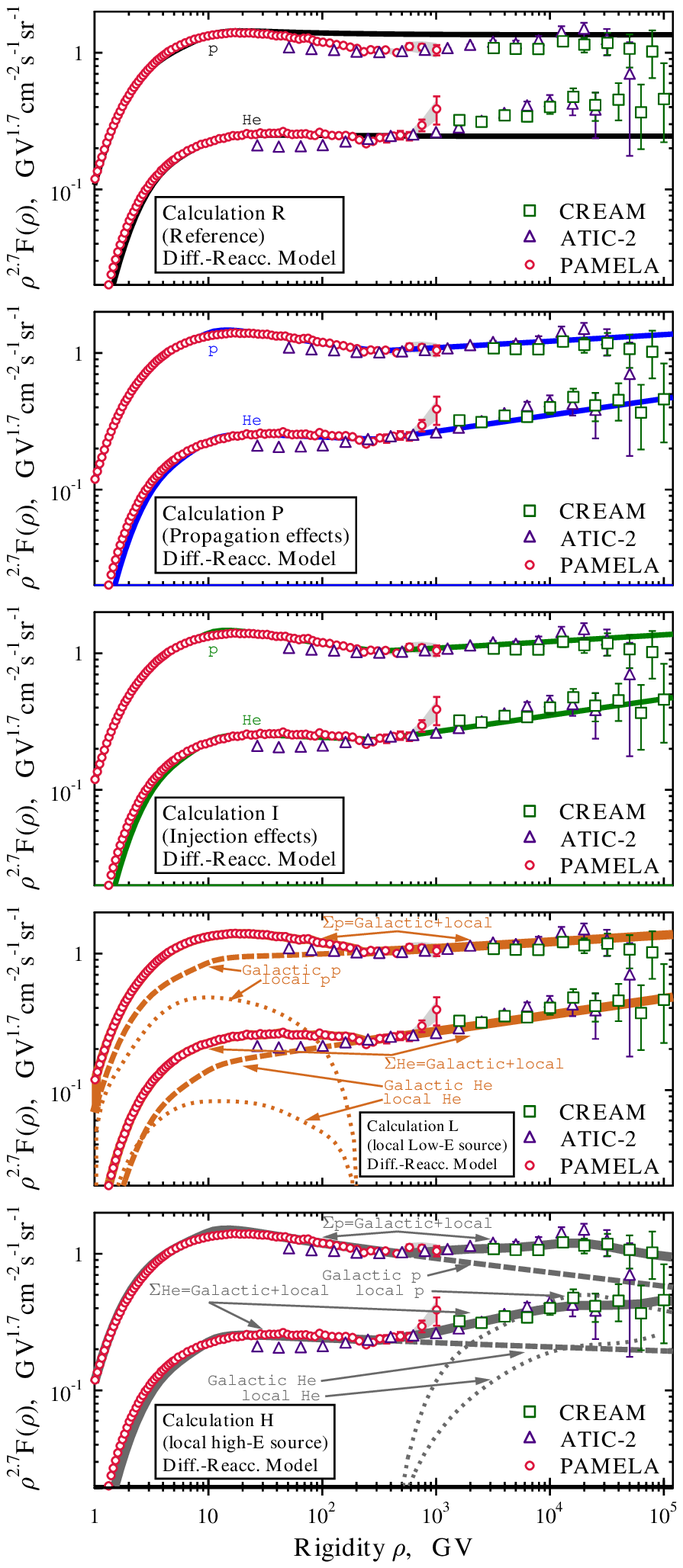}
}
\subfloat[][]{
\includegraphics[width=0.490\textwidth]{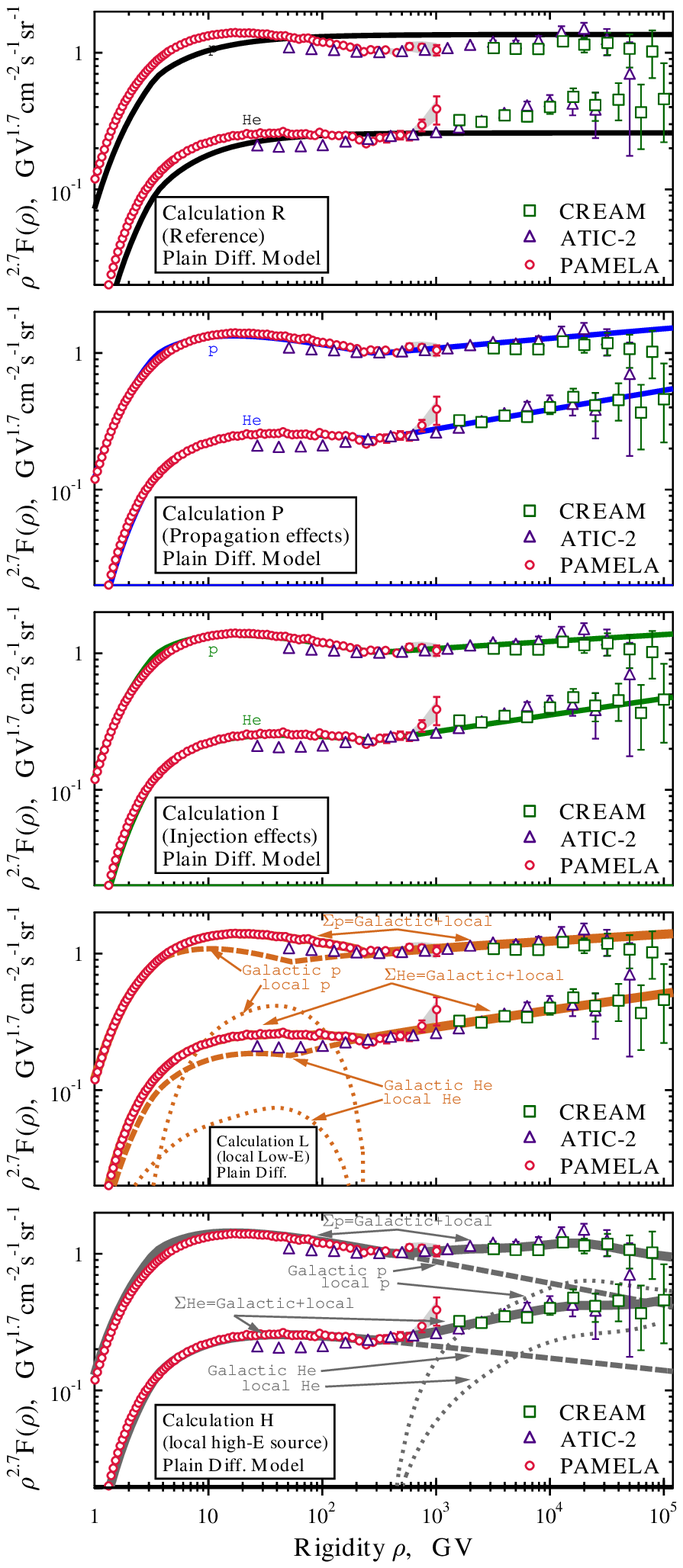}
}
}
\caption{(color in online version) Propagated CR proton and He spectra: data 
of PAMELA \citep{PAMELA2011}, ATIC-2 \citep{ATIC2008elements,Panov2009}, 
and CREAM \citep{CREAM2010break, CREAM2011pHe} together with calculation 
results. 
{\it Left}: diffusive-reacceleration model, {\it right}: plain diffusion model.
For {\it Calculations L} and {\it H}, solid lines show the net CR flux 
comprised of the Galactic source contribution (dashed lines) and a ``local'' 
source contribution (dotted lines). 
Solar modulation in all spectra is taken into account using force-field 
approximation with a modulation potential $\Phi=450$~MV. 
See the discussion in Section~\ref{results.phe}.}
\label{fig_cr_spectra}
\end{figure*}

\begin{figure}[tbh]
\center{
\subfloat[][]{
\includegraphics[width=0.490\textwidth]{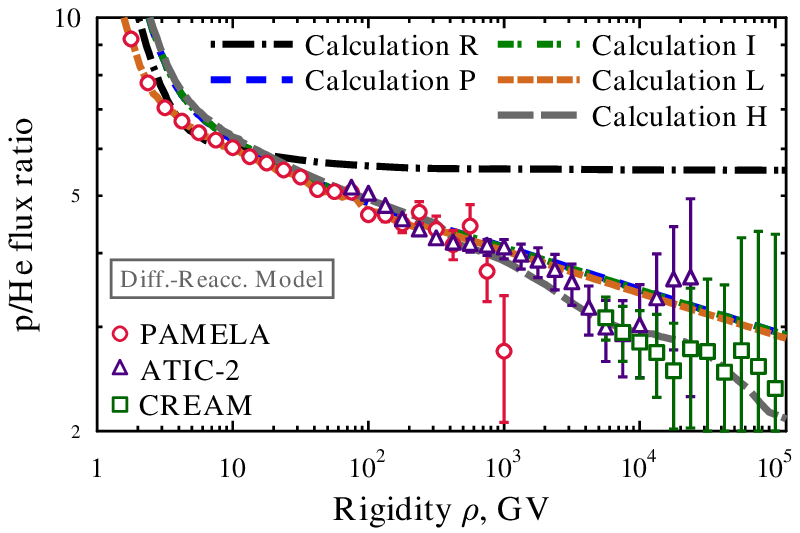}
}
\subfloat[][]{
\includegraphics[width=0.490\textwidth]{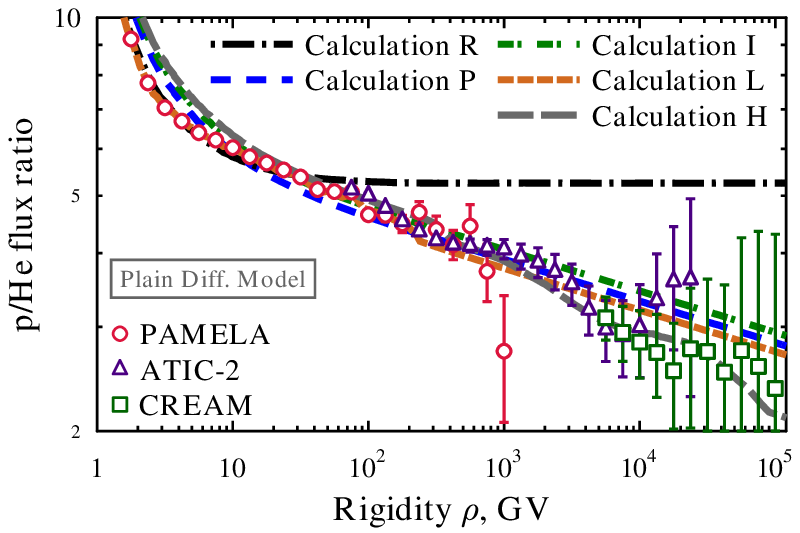}
}
}
\caption{(color in online version) Propagated CR proton to He flux ratio: data of 
PAMELA \citep{PAMELA2011}, ATIC-2 \citep{ATIC2008elements,Panov2009}, and 
CREAM \citep{CREAM2010break, CREAM2011pHe}, together with calculation results. 
{\it Left}: diffusive-reacceleration model, {\it right}: plain diffusion model.
The $p$/He points for experimental data were obtained by interpolating 
the measured fluxes of protons and He along with respective errors and 
calculating the $p$/He ratio on a grid. 
See the discussion in Section~\ref{results.phe}.}
\label{fig_p_he_ratio}
\end{figure}

Proton and He spectra calculated for the different scenarios and their ratio 
are plotted in Figures~\ref{fig_cr_spectra} and \ref{fig_p_he_ratio}. 
The bins in rigidity for protons are different from the bins for He in all 
experiments. 
Because of this, the experimental data points of PAMELA, ATIC-2, and CREAM 
shown in Figure~\ref{fig_p_he_ratio} were obtained by interpolating the 
proton and He spectra, along with their errors, and by calculating the $p$/He 
ratio on a grid. 
For simplicity, solar modulation for all spectra is taken into account using 
the force-field approximation \citep{1968ApJ...154.1011G} 
with a modulation potential $\Phi=450$~MV.

While the reference case, \calcR, provides satisfactory agreement with 
pre-PAMELA data by construction, it naturally misses 
all newly discovered features: an overall harder He spectrum, the spectral 
break at $\Rbr$, along with the dip just below $\Rbr$.
The difference between the spectrum of He and protons at all energies was 
phenomenologically included in all other calculations except \calcR, which 
is reflected in the considerably better agreement with the $p$/He ratio data in 
Figure~\ref{fig_p_he_ratio}. 

\calcP. A break in the rigidity dependence of the diffusion coefficient 
leads to a corresponding break in the CR spectrum at Earth. 
In order to match the data, we assumed the change in the value 
of $\delta$ (see Equation~(\ref{Dxx})) from $\delta_1=0.30$ to $\delta_2=0.15$ 
at $\Rdifhigh\approx \Rbr$. 
For the PD model, the index changed at
$\Rdifhigh$ from $\delta_1=0.60$ to $\delta_2=0.37$.
The difference $\delta_1 - \delta_2$ is greater in the PD than in the D-R model,
because in the latter the reacceleration process additionally softens 
the spectrum of protons below $\Rdiflow$.
Two corresponding physical quantities can be derived from these values. 
Assuming that the change in $\delta$ is caused by a difference between 
the properties of interstellar MHD turbulence on scales smaller and larger 
than a certain length scale $\lbr$, we can estimate this length to be of 
order of the gyroradius of 300~GV particles.
The gyroradius of a particle of rigidity $\rigidity$ in magnetic field $B$ is

\begin{equation}
r_g = 4 \times 10^{-2} 
\left( \frac{\rigidity}{1\,\mathrm{GV}} \right) 
\left( \frac{B}{5\,\mathrm{\mu G}} \right)^{-1}\,\mathrm{AU}.
\end{equation}

\noindent
For a characteristic interstellar magnetic field of order of a few $\mu$G, 
this implies a change in turbulence properties on length scales of the order 
of $\lbr\approx10$ AU.
If the quasilinear theory of turbulent particle diffusion applies to CR 
transport in the ISM, the value $\delta_2=0.15$ corresponds to 
turbulence spectral index, $\alpha=-2+\delta_2=-1.85$, which is harder 
than a Kolmogorov spectrum, $\alpha=-5/3$. 
Note that the direction in which the index $\alpha$ changes across the 
transition wavenumber $k=\lbr^{-1}$ is opposite to the transition of 
the turbulent cascade from the inertial to dissipative regime. 
In our case, the turbulence spectrum must harden, rather than soften, 
above $k=\lbr^{-1}$.

\calcI\ assumes a change of the power-law index of CR injection spectrum 
at $\Rinjbr=300\mathrm{~GV}$, which produces a break 
at $\Rbr \approx \Rinjbr$ in the CR spectrum at Earth. 
The dip is not produced in this calculation.

\calcL\ (dashed orange lines in Figures~\ref{fig_cr_spectra} 
and \ref{fig_p_he_ratio}) agrees with the $p$/He ratio and 
spectral break, and also reproduces the dip just below $\Rbr$. 
This is possible because of the combination of the hard spectrum from 
Galactic sources that matches the data for $\rigidity > \Rbr$ (solid orange 
lines in Figure~\ref{fig_cr_spectra}) with a local low-energy source 
having a sharp turnover just below $\Rbr$ (dotted orange lines in 
Figure~\ref{fig_cr_spectra}). 

In \calcH, the spectral break at $\Rbr$ is produced by the local source 
beginning to dominate the CR spectrum above $\Rbr$. 
We assume that the Galactic source has the same power-law index 
for $\rigidity > \Rbr$ as the low-energy CR spectrum.

The observed continuity of the $p$/He ratio and its slope at $\Rbr$ within 
statistical and systematic uncertainties is very important.  
In \scenP, this property of the $p$/He ratio comes about naturally. 
Indeed, if the injection spectrum is continuous, then protons and He 
nuclei experience the change in diffusion coefficient in the same way, and 
the $p$/He ratio is unaffected. 
However, matching this observation in the framework of a composite source 
spectrum (\scenL, \scenH, or \scenIcomp) requires an additional assumption 
of the H to He ratio to be the same at the sources producing the low-energy 
and high-energy particles. 

For the analysis of all calculations discussed above, the dip in the 
spectrum, if it is significant, may lead to important implications. 
One possible explanation for the dip may be provided in the framework 
of \scenL\ and \scenIcomp.
It can naturally appear if
the spectrum of the low-energy CR sources 
(local, as in \scenL, or Galactic, as in \scenIcomp) sharply turns 
over just below $\Rbr$, rather than continuing as a power law up 
to the knee in the CR spectrum.
Indeed, it is trivial to prove that for any two power-law spectra, their 
sum always hardens with energy. 
Thus, for the softening below $\Rbr$ to occur, the low-energy source 
spectrum may not be a pure power law; the sharpness of the dip suggests 
that it must steeply turn over below $\Rbr$, where the dip occurs. 
The dip may also be explained in the framework of \scenP, if a corresponding 
dip in the spectrum of MHD turbulence responsible for 
CR confinement in the Galaxy is assumed. 
It is not possible to explain the dip with \scenH\ because the low-energy 
source is assumed to have a power-law spectral shape extending all the way 
to the knee.

\subsection{Electrons}\label{results.electrons}

The CR electron spectrum in this problem is connected
to the proton spectrum because (1) electrons propagate
in the Galaxy in the same magnetic fields as nuclei, and (2) some, if not
all, CR electrons are produced by the same sources as nuclei.

\begin{figure*}[t]
\center{
\subfloat[][]{
\includegraphics[width=0.490\textwidth]{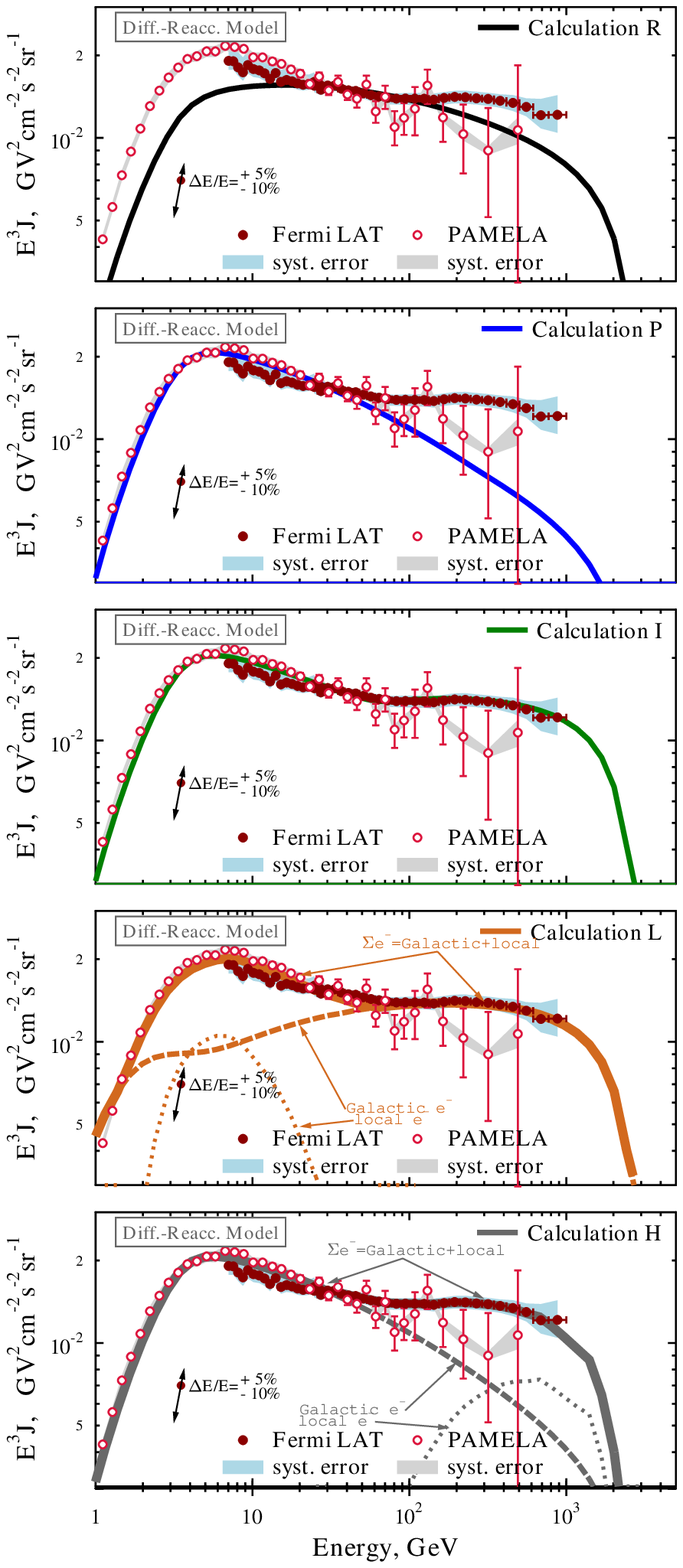}
}
\subfloat[][]{
\includegraphics[width=0.490\textwidth]{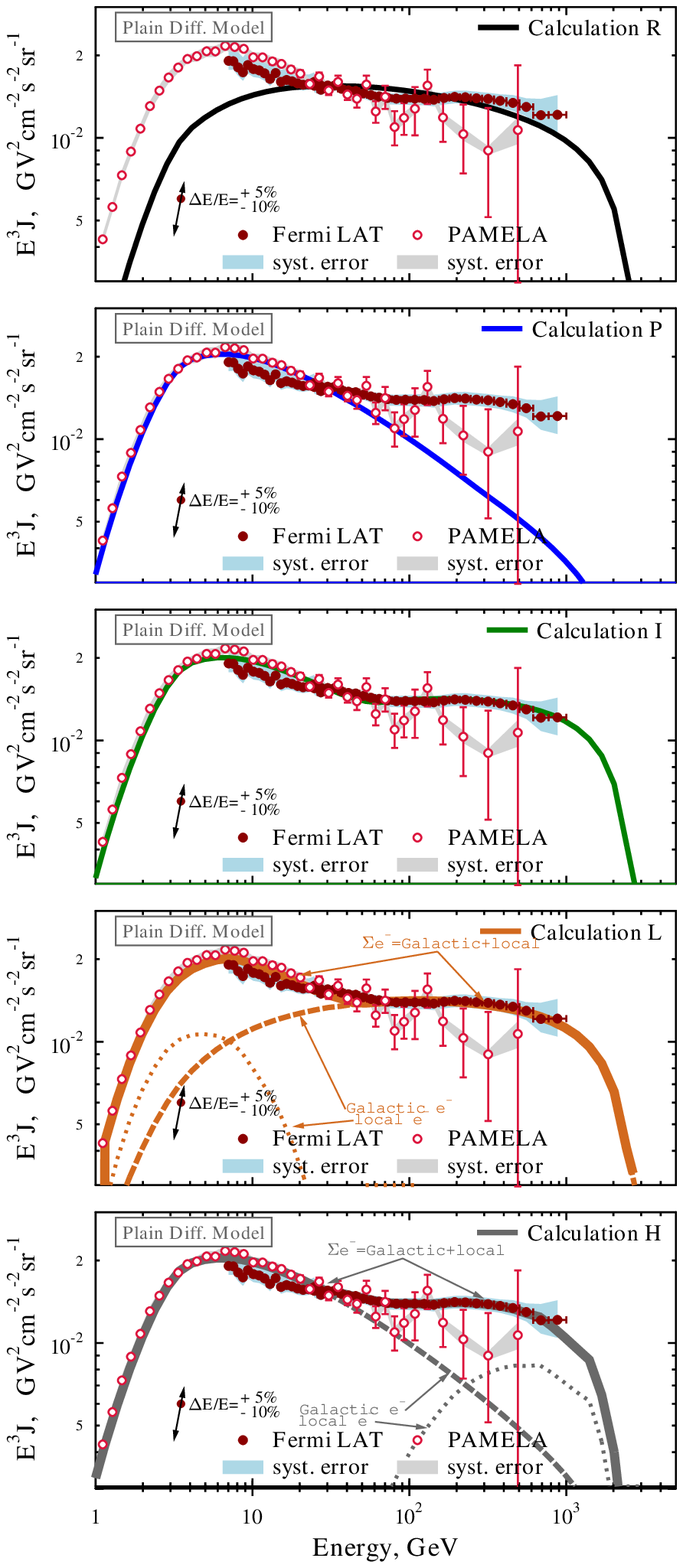}
}
}
\caption{
(color in online version) Propagated CR electron spectra: models and data.
{\it Left}: diffusive-reacceleration model, {\it right}: plain diffusion model.
The data are from \cite{FermiElectrons2010} (\fermilat) and \cite{PAMELA2011electrons}
(PAMELA). 
See the discussion in Section~\ref{results.electrons}}
\label{fig_electrons}
\end{figure*}

\begin{figure}[tbh]
\center{
\subfloat[][]{
\includegraphics[width=0.490\textwidth]{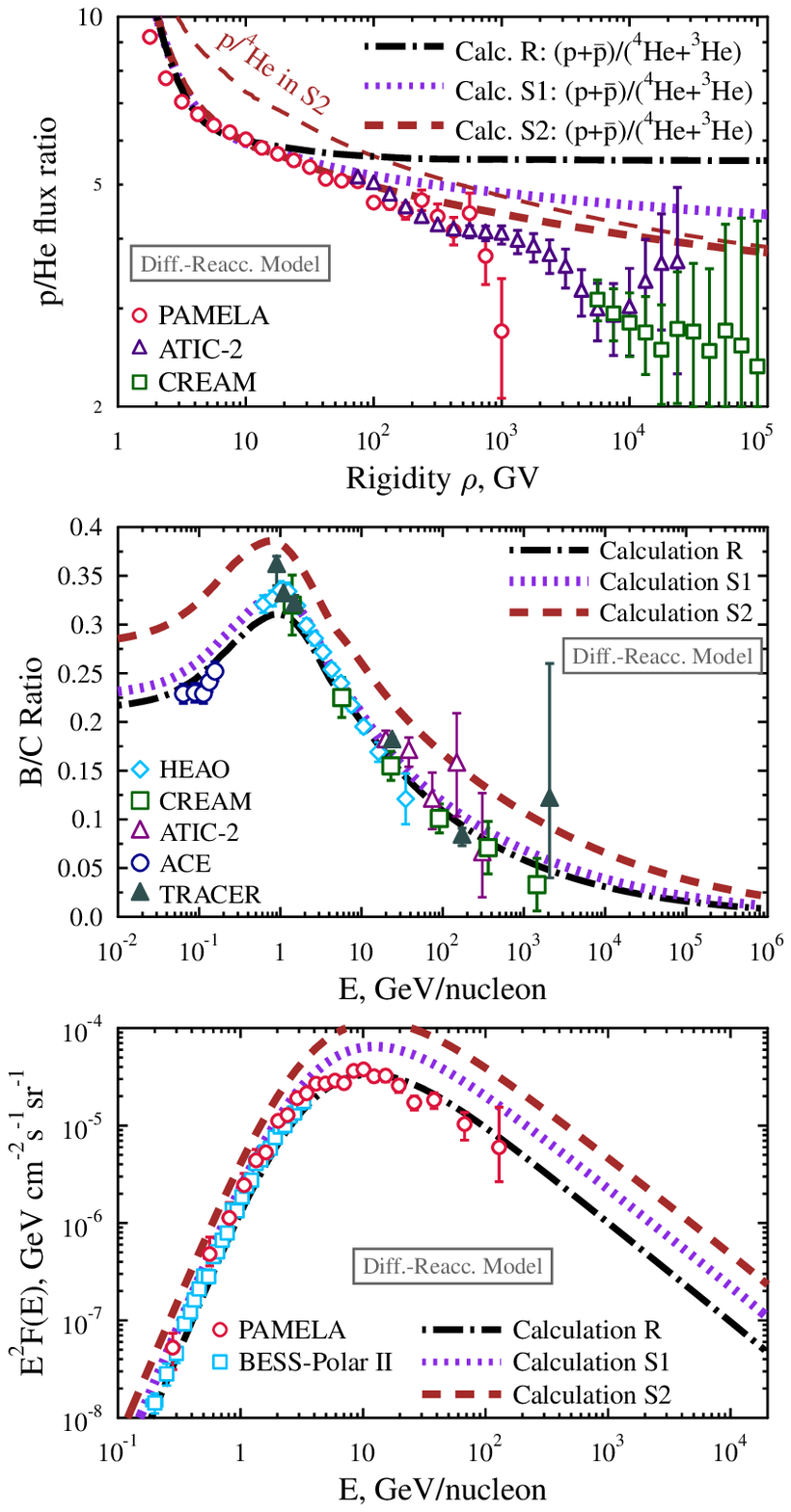}
}
\subfloat[][]{
\includegraphics[width=0.490\textwidth]{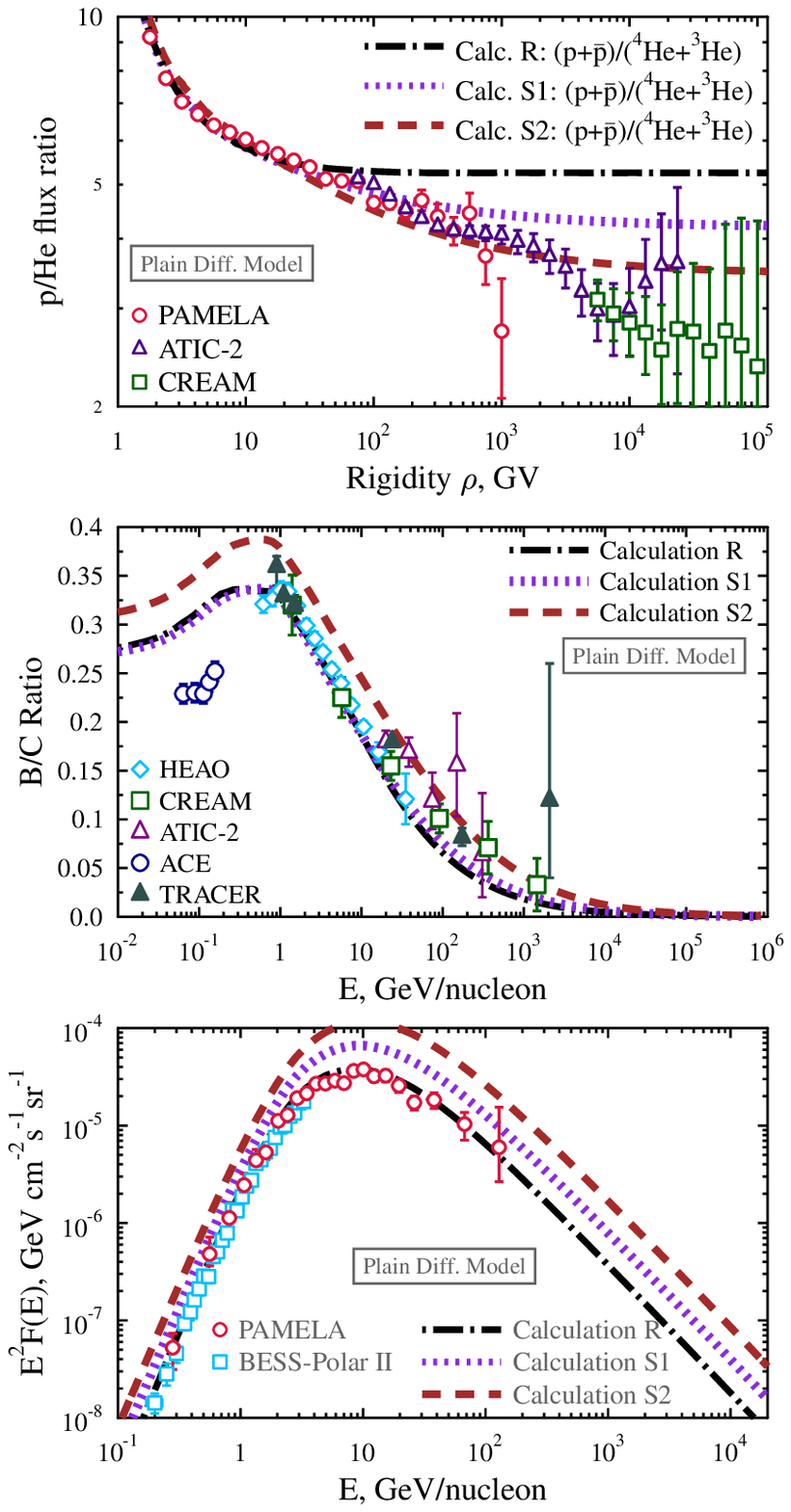}
}}
\caption{(color in online version) CR proton to He, B/C ratio and antiproton 
flux (sources of data as in Figures~\ref{fig_p_he_ratio}, 
\ref{fig_b_c_ratio}, \ref{fig_pbar}), together with calculation results. 
{\it Left}: diffusive-reacceleration model, {\it right}: plain diffusion model.
\calcS$_1$ and \calcS$_2$ use the same spallation cross sections and $z_H/D$ ratio 
as those of \cite{BA2011a}. 
In \calcS$_2$, the gas density is increased by a factor of 2 with respect 
to \calcS$_1$. 
Results of \calcS$_1$ are consistent with the B/C ratio data, but slightly 
overpredict \pbar\ measurements.
However, note that the effect of spallation is not sufficient to make the 
slope of the $p$/He ratio agree with the PAMELA data. 
In \calcS$_2$, spallation is stronger due to increased grammage, and the $p$/He 
ratio is reproduced better, but the B/C ratio and \pbar\ flux are 
overpredicted. }
\label{fig_spallation}
\end{figure}

\begin{figure}[tbh]
\center{
\subfloat[][]{
\includegraphics[width=0.490\textwidth]{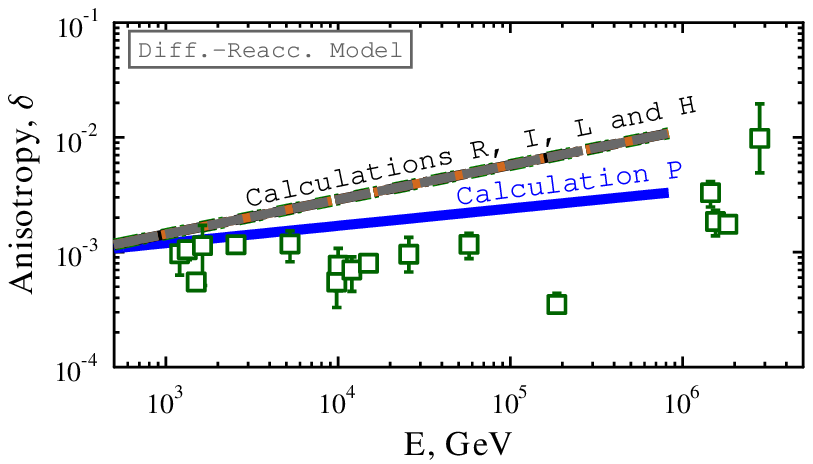}
}
\subfloat[][]{
\includegraphics[width=0.490\textwidth]{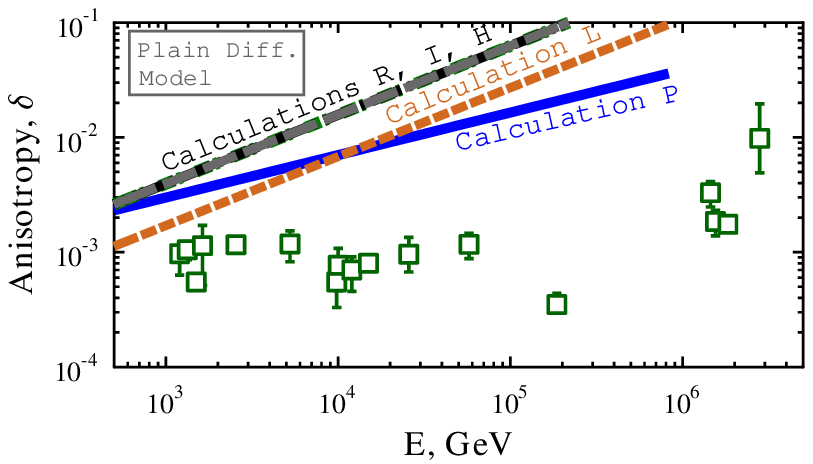}
}
}
\caption{(color in online version) CR flux anisotropy: data from 
\cite{PJSS2006}, together with calculation results. 
The anisotropy was calculated as the ratio of diffusive flux in the 
radial direction to the isotropic flux at the corresponding energy. 
{\it Left}: diffusive-reacceleration model, {\it right}: plain diffusion model.
{\it Calculations R, S, I, L} and {\it H} all predict the same value of 
anisotropy, and their respective lines overlap. 
The result of \calcP\ is different due to a different form of the CR 
diffusion coefficient. See the discussion in Section~\ref{results.anisotropy}}
\label{fig_anisotropy}
\end{figure}

\begin{figure}[tbh]
\center{
\subfloat[][]{
\includegraphics[width=0.490\textwidth]{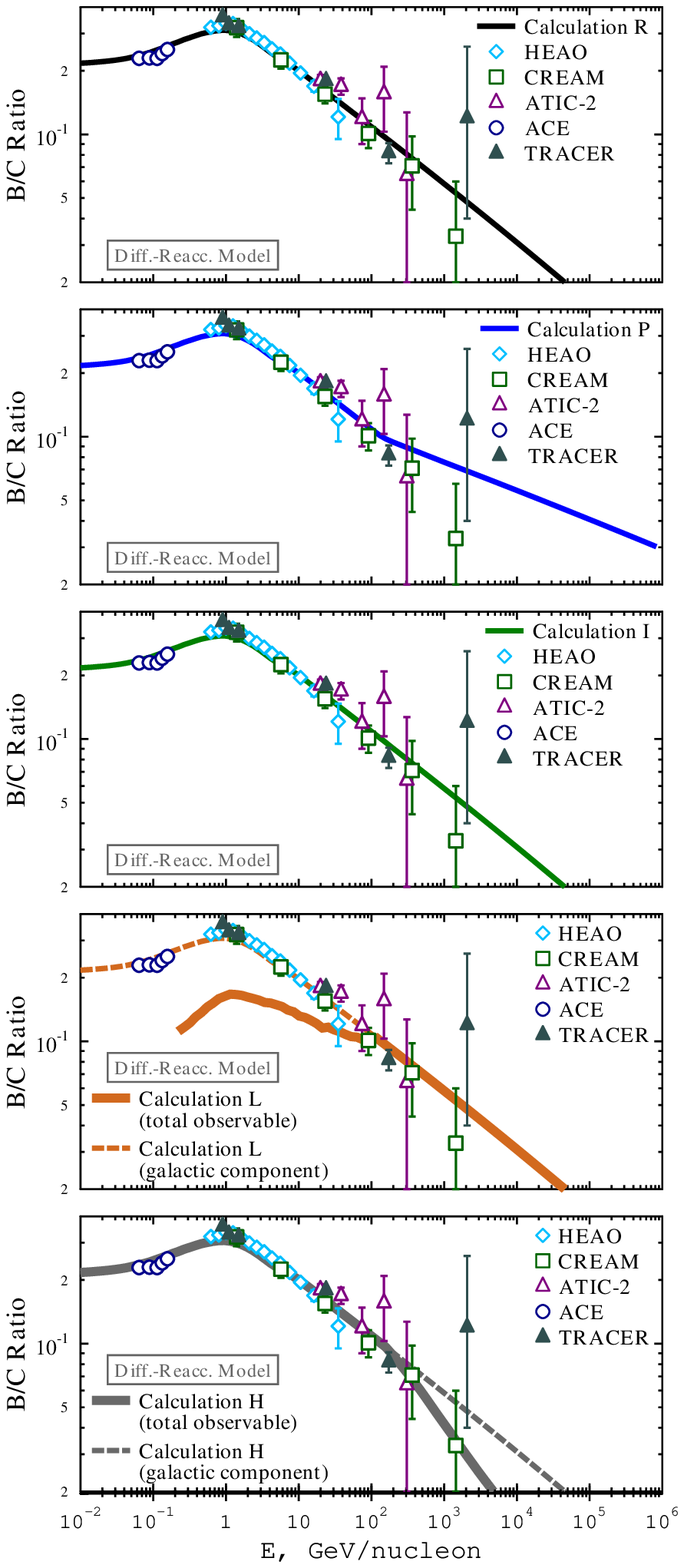}
}
\subfloat[][]{
\includegraphics[width=0.490\textwidth]{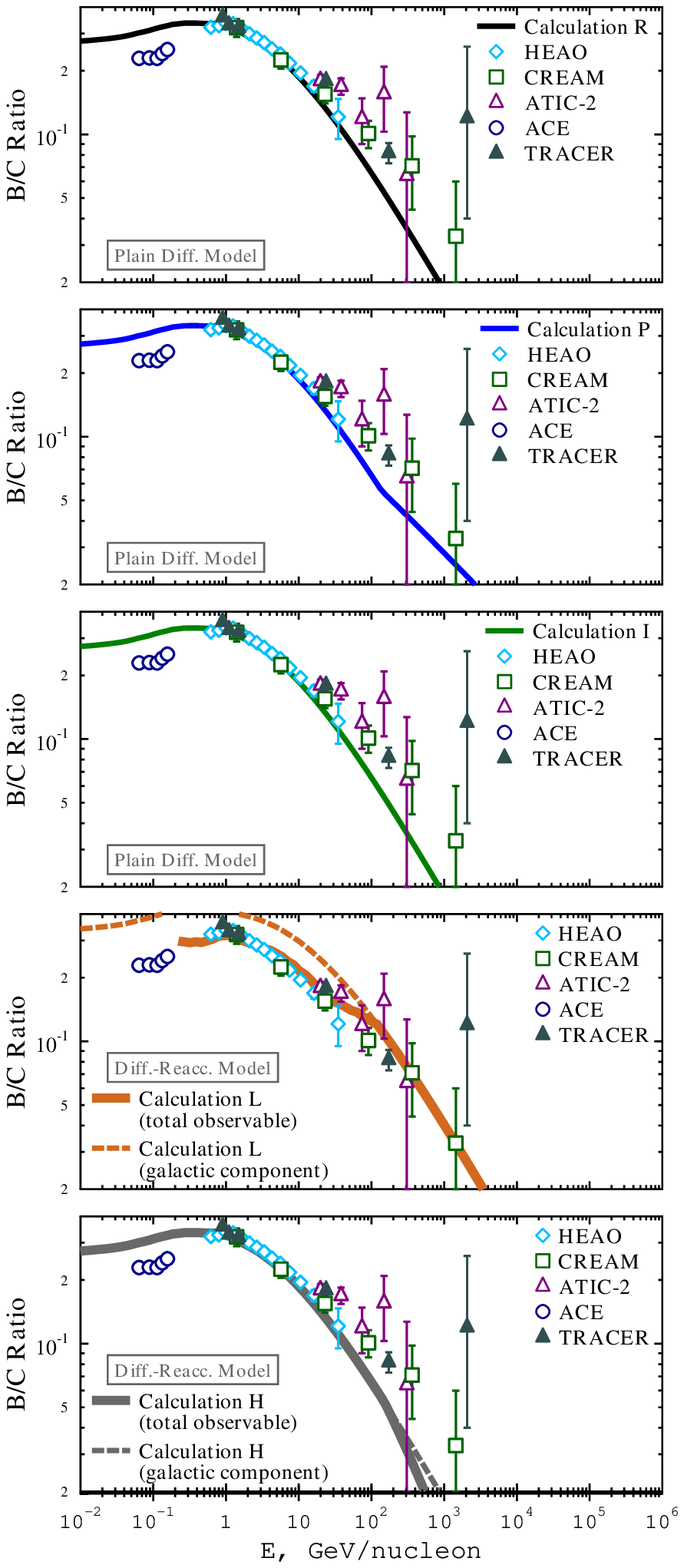}
}
}
\caption{(color in online version) CR boron-to-carbon flux ratio. 
{\it Left}: diffusive-reacceleration model, {\it right}: plain diffusion model.
Data: \cite{ACEBC} (ACE), \cite{HEAO3} (HEAO-3), \cite{CREAMBC} (CREAM),
\cite{ATIC2BC} (ATIC-2) and \cite{TRACER2011} (TRACER).
For \calcL\ and \calcH, dashed lines show the ratio of just the Galactic 
source, while solid lines show the B/C ratio including the contribution of 
the ``local'' source component. 
Additional discussion in Section~\ref{results.bc}}
\label{fig_b_c_ratio}
\end{figure}

\begin{figure}[tbh]
\center{
\subfloat[][]{
\includegraphics[width=0.490\textwidth]{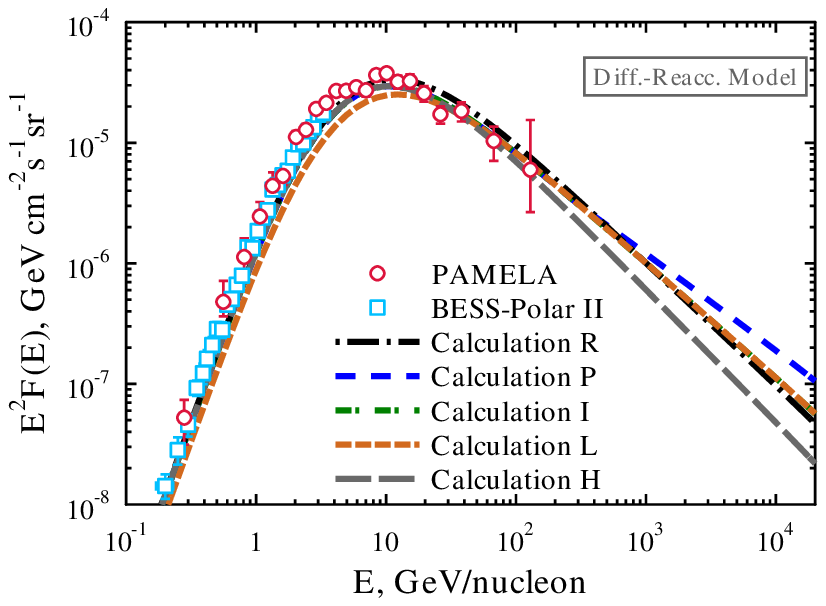}
}
\subfloat[][]{
\includegraphics[width=0.490\textwidth]{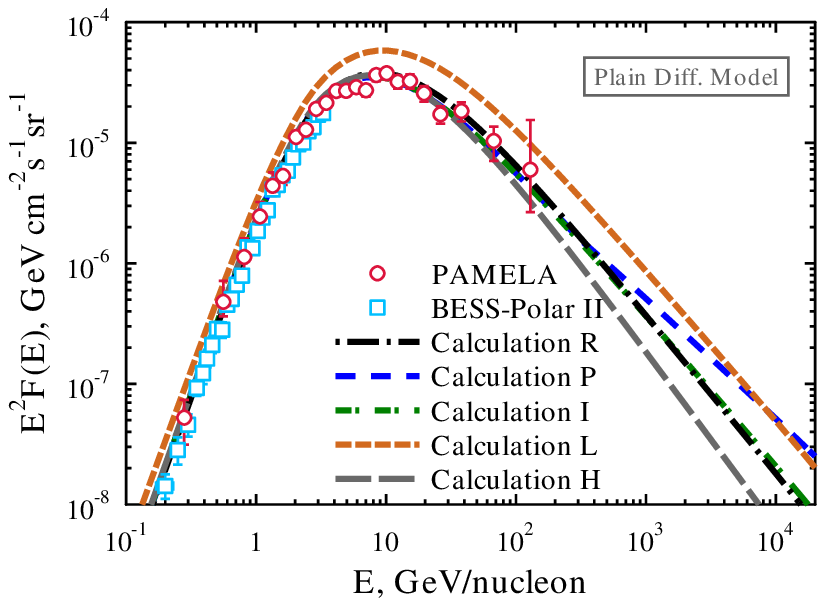}
}
}
\caption{(color in online version) CR antiprotons: data from 
\cite{PAMELApbar} (PAMELA) and \cite{BESS2011} (BESS) together with calculation results. 
{\it Left}: diffusive-reacceleration model, {\it right}: plain diffusion model.
See the discussion in Section~\ref{results.pbar}}
\label{fig_pbar}
\end{figure}

\begin{figure}[tbh]
\center{
\subfloat[][]{
\includegraphics[width=0.490\textwidth]{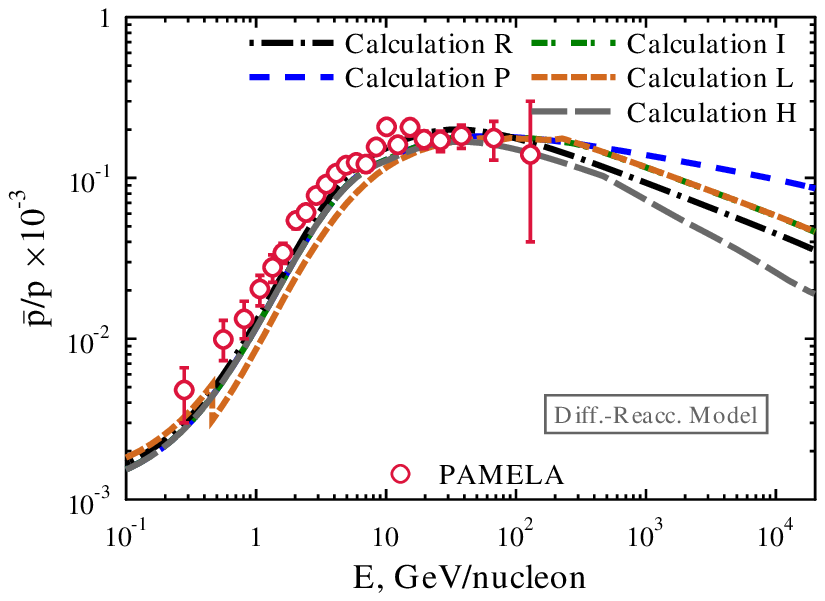}
}
\subfloat[][]{
\includegraphics[width=0.490\textwidth]{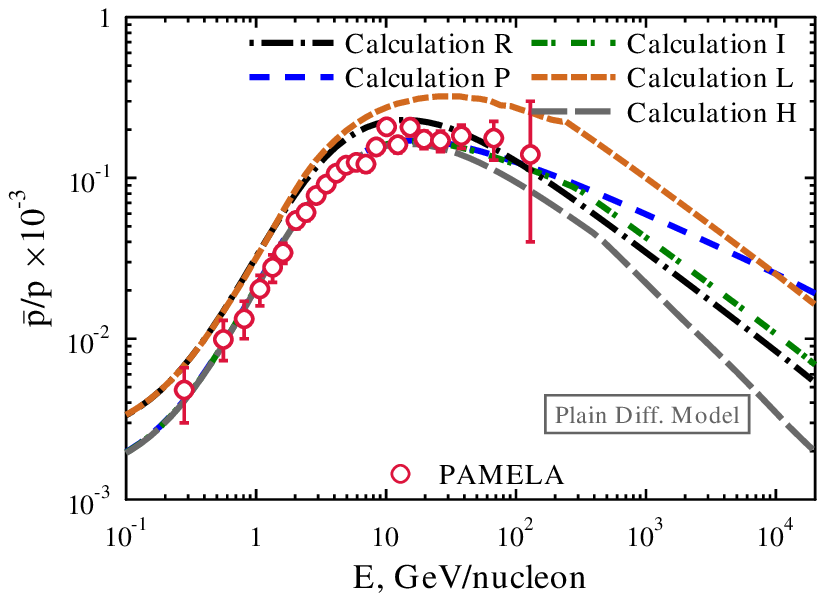}
}
}
\caption{(color in online version) CR antiproton to proton ratio: data 
from \cite{PAMELApbar} (PAMELA) together with calculation results. 
{\it Left}: diffusive-reacceleration model, {\it right}: plain diffusion model.
See the discussion in Section~\ref{results.pbar}}
\label{fig_pbar_p}
\end{figure}

\begin{figure*}[t]
\center{
\subfloat[][]{
\includegraphics[width=0.490\textwidth]{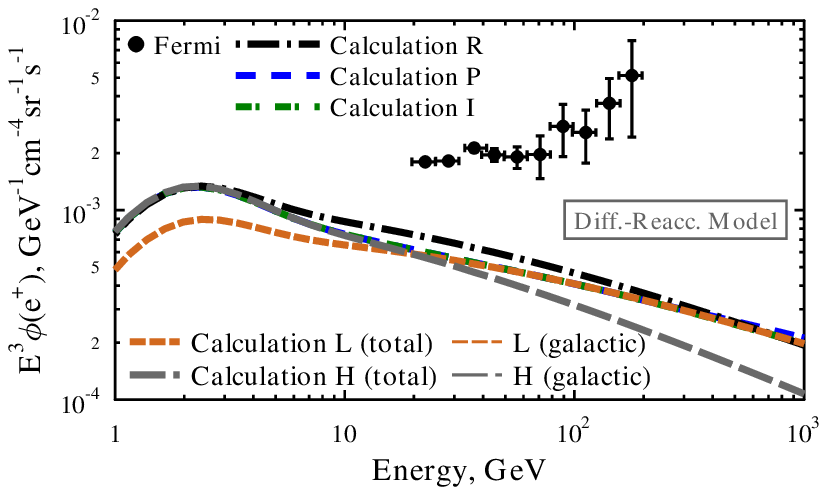}
}
\subfloat[][]{
\includegraphics[width=0.490\textwidth]{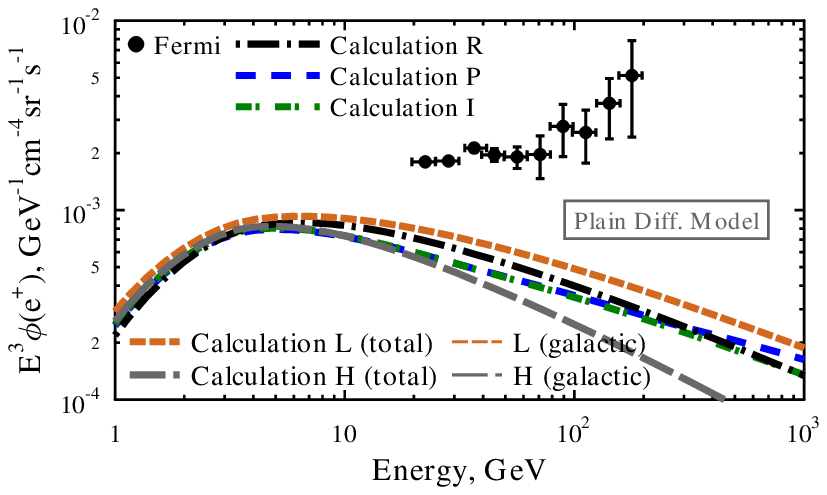}
}
}
\caption{
(color in online version) Positron flux: models and data.
{\it Left}: diffusive-reacceleration model, {\it right}: plain diffusion model.
The data are from \cite{Fermi2012positrons}
(\fermilat). 
See the discussion in Section~\ref{results.positrons}}
\label{fig_positrons}
\end{figure*}

\begin{figure*}[t]
\center{
\subfloat[][]{
\includegraphics[width=0.490\textwidth]{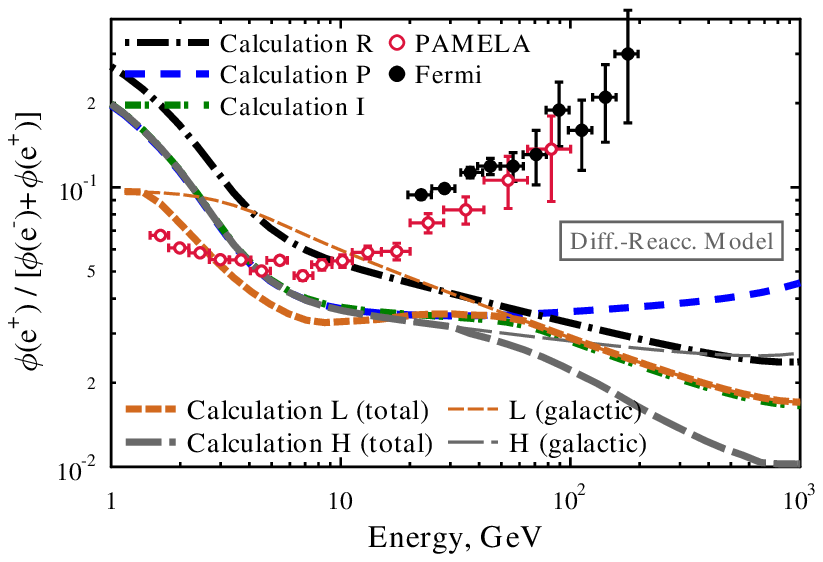}
}
\subfloat[][]{
\includegraphics[width=0.490\textwidth]{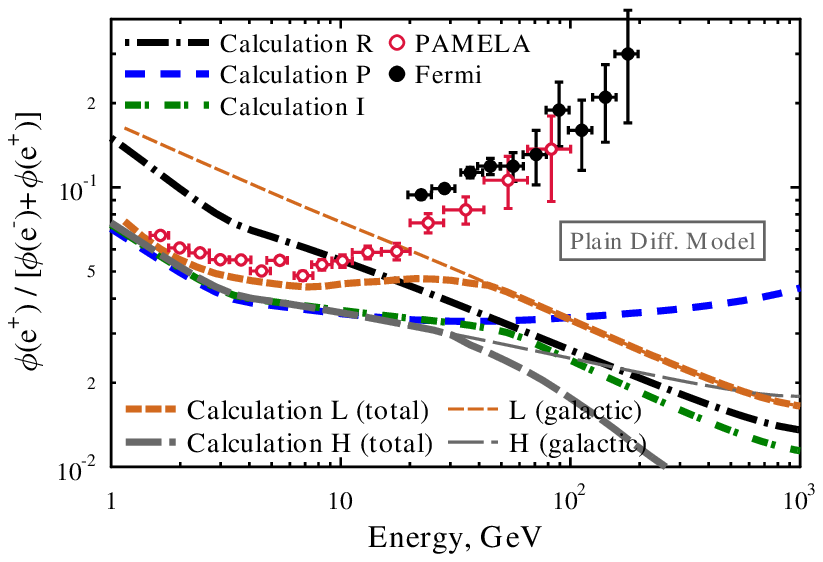}
}
}
\caption{
(color in online version) Positron fraction: models and data.
{\it Left}: diffusive-reacceleration model, {\it right}: plain diffusion model.
The data are from \cite{PAMELA2009positrons}
(PAMELA) and \cite{Fermi2012positrons}
(\fermilat). 
See the discussion in Section~\ref{results.positrons}}
\label{fig_positron_fraction}
\end{figure*}

\begin{figure*}[t]
\center{
\subfloat[][]{
\includegraphics[width=0.490\textwidth]{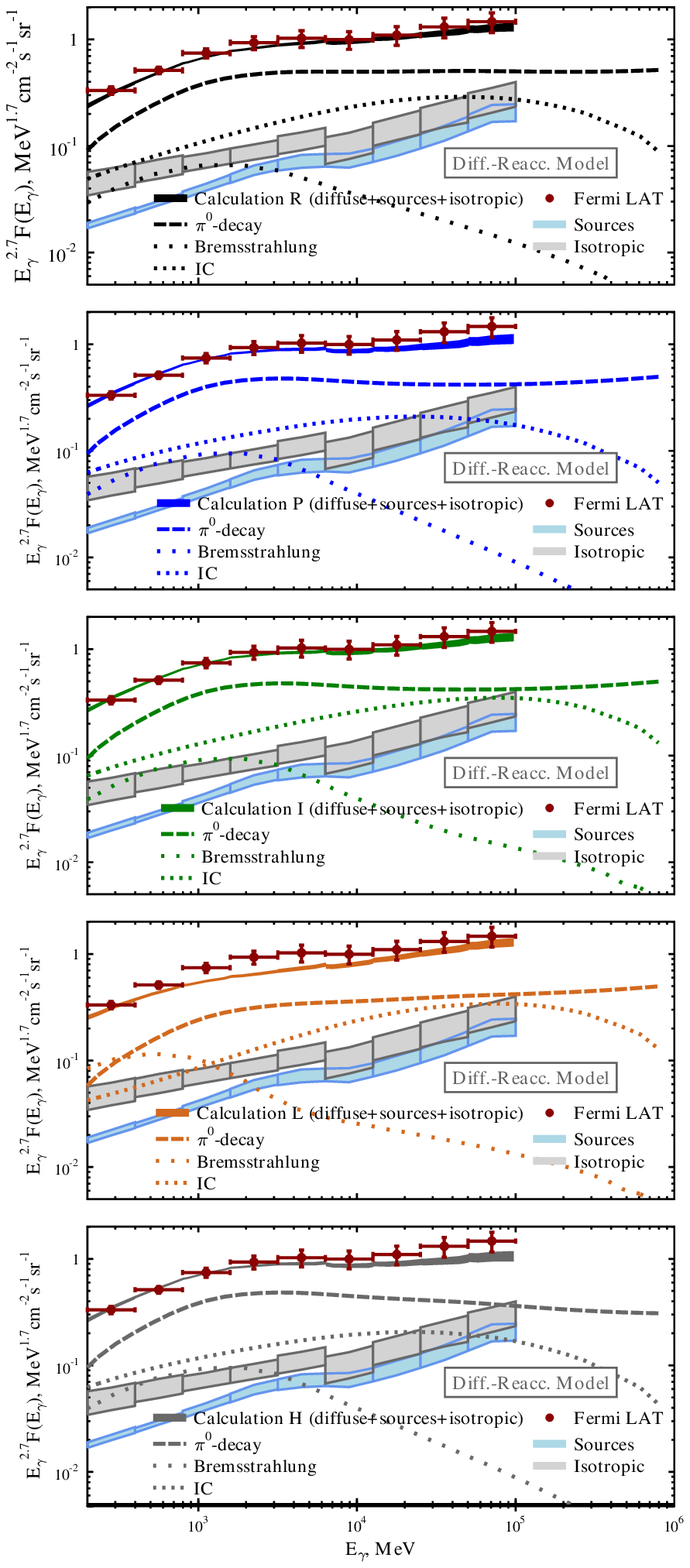}
}
\subfloat[][]{
\includegraphics[width=0.490\textwidth]{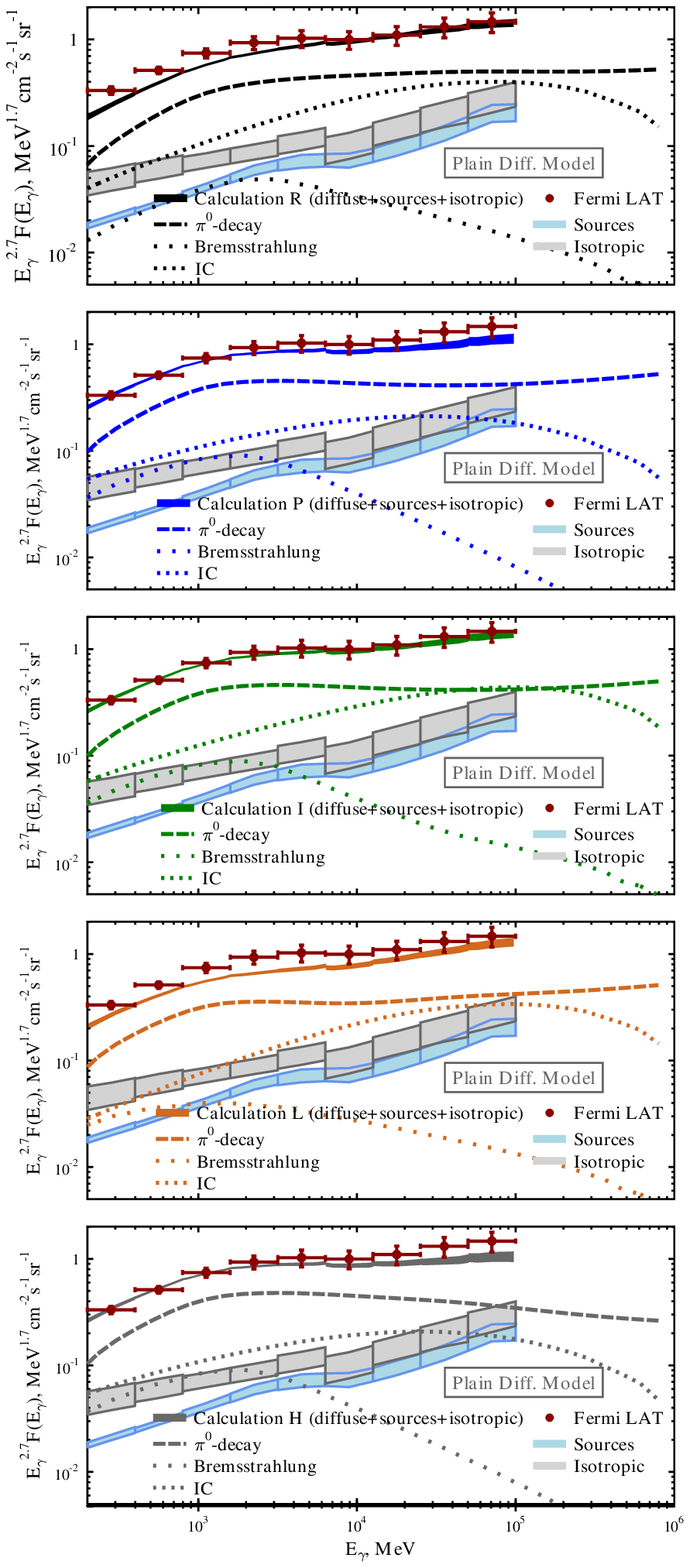}
}
}
\caption{
(color in online version) Diffuse \gray\ emission from intermediate 
Galactic latitudes: data from \cite{FermiIsotropic2010} (\fermilat) together with
calculation results. 
{\it Left}: diffusive-reacceleration model, {\it right}: plain diffusion model.
The data are the diffuse \gray\ intensities averaged over all Galactic 
longitudes and intermediate Galactic latitudes $10^{\circ} < |b| < 20^{\circ}$, 
as reported by \cite{FermiIsotropic2010} (available in the online 
supplementary material to the article). 
See the discussion in Section~\ref{results.gamma}}
\label{fig_gamma_ray_midlat}
\end{figure*}

As Figure~\ref{fig_electrons}
shows, the propagated electron spectrum in \calcR\ 
does not fit the observations of the \fermilat\ \citep{FermiElectrons2010}
or PAMELA \citep{PAMELA2011electrons}. Indeed, the observed spectrum appears
convex (i.e., hardening with increasing particle energy) from $\approx$10 to $10^3$~GeV,
whereas the calculated spectrum in this energy range is concave. The concavity
is caused by energy losses on ionization at low energies, and synchrotron losses
at high energies.

In \calcP, despite a break in the diffusion coefficient, the electron spectrum 
at Earth does not fit the high-energy data and is not convex (Figure~\ref{fig_electrons}). 
This is because synchrotron energy losses above 100~GeV oppose the effect
of the diffusion coefficient break, and cause the spectral softening.

In the injection effects Scenario, represented by \calcI, it is possible
to modify the source spectrum of electrons in order to fit the data. Indeed, since
the nucleon injection spectrum has a break at $\Rbr$, it is 
natural to assume that the electron injection spectrum may have a similar feature. 
This argument holds both in the 
source with a spectral break interpretation (\scenIbr) and the
composite spectrum interpretation (\scenIcomp) of \calcI. 
Moreover, the energy and magnitude
of the break do not have to be the same for electrons and protons because
the electron to proton ratio may vary with energy and with source type.
As Figure~\ref{fig_electrons} shows, one can achieve agreement with the data
by assuming that the electron source spectrum has a break (hardening) at $R=70$~GV
(61~GV in the PD model),
with the index changing from $-2.70$ to $-2.33$ (or $-2.67$ to $-2.24$ for PD). 
Note that one cannot
justify such a break in any other scenario, unless CR electrons and nuclei are assumed
to be produced by different sources.

In \calcL, we assumed that
above the break ($\gtrsim 100$~GV for electrons) the particles are produced by a hard 
Galactic source, and that below the break the electron flux is dominated by an unknown 
{\it low}-energy local source. Likewise,
in \calcH, an unknown {\it high}-energy source of CR electrons was assumed. 
In both cases, the flux of the local
source was calculated as the difference between the observed and the calculated Galactic source
fluxes. Therefore, the total spectrum agrees with the data at all energies. 
Note that only the Galactic source
flux was used in the calculation of secondary lepton production and \gray\ emission.

\subsection{Anisotropy}\label{results.anisotropy}

For all scenarios, we calculated the anisotropy of the high-energy CR flux 
at the location of the Sun due to diffusive escape of CRs from the Galaxy. 
The results are presented in Figure~\ref{fig_anisotropy}, along with data. 
References to individual experiments may be found in \citet{PJSS2006}; see 
also \citet{SMP2007} for a color version of the plot.
The anisotropy, dominated by the radial component, is highly sensitive to 
the choice of the diffusion coefficient and the spatial distribution of 
CR sources. 

Our calculation ignores the effect of nearby CR sources, which may be 
significant \citep[][]{PJSS2006, BA2011b}, but is not very well defined 
and depends on the assumed distances to the sources and their ages. 
However, if the diffusive component of the anisotropy dominates in a 
certain energy range, two conclusions may be drawn from this plot.
The first one is that \scenP\ can be distinguished from the others 
with improved CR proton anisotropy data.
The second point illustrated by our calculation is that the diffusion 
regime of \calcP\ with $D\propto \rigidity^{\delta_2}$ and $\delta_2=0.15$ agrees 
with the available data better than $\delta_2=0.30$. 
The PD model, due to a harder dependence of the diffusion coefficient on rigidity, 
predicts a higher degree of anisotropy, which disagrees with the data more 
than the other calculations. 

In the case of \calcH, the plotted lines correspond to only the Galactic 
source, while the direction and magnitude of the local source flux 
anisotropy above $\Rbr$ are unknown. Depending on the location
and proximity of the local source, the direction of the overall
CR drift can be changed substantially, because the local source flux
above $10^{3}$~GV is comparable to the Galactic flux (see
the bottom plots in Figure~\ref{fig_cr_spectra}). Since the estimate 
of the distance to the local source is beyond the scope of this paper,
we do not provide quantitative predictions of anisotropy in this case.

\subsection{Boron to Carbon Ratio}\label{results.bc}

The B/C ratio for 
all scenarios discussed in the paper is shown in Figure~\ref{fig_b_c_ratio}. 
Predictions of \calcR\ and \calcI\ coincide at all energies, while \calcP\ 
predicts a larger B/C ratio for $\rigidity > \Rbr$, which is a consequence 
of the smaller diffusion coefficient in \calcP. 
In the case of \calcL\ and \calcH, the results include the effect of the 
local CR source. 
CR boron is produced by fragmentation of heavier elements and decay 
of $^{10}$Be. 
If the local source is very nearby, its flux should contain no boron.
However, the abundance of (primary) 
carbon should be close to the interstellar value, which 
results in a lower B/C ratio in the net flux than without the local source.

To find the B/C ratio for \calcH\ and \calcL, we assume that the local 
source produces no boron and assume that the flux of local source carbon is 
proportional to that of He, with the same carbon and He abundance as in 
the Galactic source. 
\calcH\ predicts a lower B/C for $\rigidity > \Rbr$ because the local 
high-energy source supplies primary carbon, but not secondary boron, at these 
energies. 
\calcL\ could not be tuned to reproduce the B/C ratio in the D-R model
(see Section~\ref{models}) However, in the PD model, the only calculation
fitting the B/C ratio below 1~GeV~nucleon$^{-1}$ is \calcL\ due to the
reduction of the B/C ratio by the local source.

Experimental data at low energies (below 1~GeV~nucleon$^{-1}$) were collected 
by {\it ACE} \citep[][]{ACEBC}, and for high energies by {\it HEAO-3} \citep[][]{HEAO3}, 
CREAM \citep[][]{CREAMBC}, ATIC-2 \citep[][]{ATIC2BC}, 
and TRACER \citep[][]{TRACER2011}.
The uncertainties in the data are still too large to rule out any of the 
scenarios considered in this paper, but data collected by future experiments, 
such as AMS-2, may be more constraining. 

\subsection{Antiproton Flux and \pbarit/$p$ Ratio}\label{results.pbar}

The antiproton flux, another probe of CR propagation, is plotted in 
Figure~\ref{fig_pbar}, and \pbar/$p$ ratio in Figure~\ref{fig_pbar_p}, together
with the PAMELA data from \cite{PAMELApbar} and the 
\mbox{BESS-Polar~II} data from \cite{BESS2011}. 
\calcs{R, P, I, {\rm and} H} are in good agreement with data below 100~GeV.
Differences between all calculations are apparent above $\sim$1~TeV, but 
no data are currently available. 

\calcL, the only case of the PD model where the low-energy B/C data
are reproduced,
predicts a factor of $\sim2$ excess of \pbar\ below 100~GeV, due to a larger particle confinement time.
In \calcs{L {\rm and} H}, the local source was assumed to be completely 
devoid of primary or secondary antiprotons. 
More accurate data covering a larger energy range may help eliminate 
some of the scenarios, and the AMS-2 mission may provide these data.

\subsection{Positrons}\label{results.positrons}

Figure~\ref{fig_positrons} shows the calculated positron flux,
and Figure~\ref{fig_positron_fraction} --- the positron fraction ($e^+/(e^-+e^+)$).
Our model does not include a source of primary CR positrons;
the $e^+$ particles in all our calculations are produced in inelastic collisions
of other CR species.

The positron flux measured by the \fermilat\ \citep{Fermi2012positrons} is
significantly greater than the prediction of all our calculations
(Figure~\ref{fig_positrons}).
The only scenario that could explain the discrepancy is 
\scenH\ in which the local high-energy CR source produces {\it primary}
positrons. 

The positron fraction (Figure~\ref{fig_positron_fraction}) provides additional
evidence that the calculations predict insufficient flux of positrons at high
energies. However, below a few GeV, the positron fraction in the D-R
model is overpredicted.


\subsection{Diffuse \gRay\ Emission}\label{results.gamma}

Predictions of the \gray\ emission at intermediate Galactic 
latitudes ($10^{\circ} < |b| < 20^{\circ}$) are plotted in 
Figure~\ref{fig_gamma_ray_midlat} along with the data reported 
by \citet[][see the online supplementary material]{FermiIsotropic2010}.
Following \citet{FermiIsotropic2010}, 
the flux of the inverse Compton (IC) component was 
increased by a factor of two to obtain a good fit to the data. 
Our calculations include \gray\ emission produced by hadronic and leptonic 
components of CRs (i.e., $\pi^0$-decay, IC, and bremsstrahlung channels), as 
well as point sources, and the isotropic extragalactic emission. 
The relative differences in the {\it total} \gray\ flux between the 
considered scenarios are quite small and are considerably smaller than if 
only the $\pi^0$-decay channel is considered \citep[as in][]{Donato2011}. 

Predictions of all calculations, except \calcL, agree with 
the published \fermilat\ data, within the uncertainty band. 
\calcL\ predicts a slightly lower \gray\ emission below 10~GeV.
Note that even the reference \calcR, which does not agree with the 
PAMELA data, satisfactorily reproduces the \gray\ data. 

For all scenarios we calculated the \gray\ spectrum up to 1~TeV, but the 
\fermilat\ team has not published on the data above 100~GeV so far. 
At these energies,
the softer spectrum of protons (above $\Rbr$) from Galactic sources in \calcH\ 
produces less pions resulting in a smaller flux of pionic \gray{s}
compared to other scenarios.
However, the contributions of comparable IC component and isotropic emission 
in the range 100~GeV -- 1~TeV are not affected by the proton spectrum. 
Therefore, even at these energies the difference between 
the {\it total} \gray\ emission in \calcH\ and other calculations is 
significantly smaller than the difference in the $\pi^0$-decay channel alone.
Unsurprisingly, \calcP\ cannot be distinguished from \calcI\ using 
the \gray\ data alone, because calculations for both scenarios result in 
nearly the same spectrum of CR protons, even though it is achieved via 
different mechanisms.

As the {\it Fermi} mission continues, the statistical uncertainty will be reduced 
as data accumulates, and systematic errors are likely to be brought down by 
improved data analysis. 
It should be noted, however, that the analysis of the diffuse \gray\ emission 
is complicated by many factors, including the uncertainty in the spatial 
distribution for the CR sources and the loosely constrained spectrum of 
CR electrons over the Galaxy 
responsible for IC emission that dominates 
high-energy \gray{s}.

\section{Summary}

We have presented scenarios reproducing the spectral features in CR 
proton and He spectra (the $p$/He ratio dependence on energy, the dip, 
and spectral break) observed by ATIC-2, CREAM, and PAMELA. 
For each scenario, we performed CR propagation calculations in the framework of 
the D-R model (except \scenL), using the GALPROP code. 
Differences between scenarios are reflected in the CR anisotropy and 
fluxes of secondary CR species: the B/C ratio at high energies, the 
antiproton flux and antiproton to proton ratio, as well as the diffuse 
Galactic \gray\ emission. 
We find the following:

(1) He spallation (\hypS) may be partially responsible for making the spectrum of He and heavier nuclei harder than protons. 
However, a significantly increased grammage traversed by CRs in the Galaxy 
is required to explain the $p$/He observations with spallation alone. 
This makes it problematic to match stable secondary CR isotope 
observations (B/C and antiprotons).

(2) Electron spectrum can be reproduced in \scenIbr\ (break in the injection spectrum)
or \scenIcomp\ (composite Galactic source)
only if the break in the electron injection spectrum is stronger, 
and occurs at a lower rigidity, than in the proton spectrum. A break in the
diffusion coefficient (\scenP) cannot simultaneously explain the 
concavity of the observed proton and electron spectra.

(3) Experimental uncertainty in the data on high-energy B/C ratio does 
not allow us to rigorously reject any of the scenarios for the origin 
of the spectral break. However, more accurate measurements of high-energy 
B/C, expected from planned CR experiments, may be used for model rejection.

(4) In the D-R model, low-energy B/C data are consistent with any scenario except 
\scenL\ (low-energy local source). In the PD model, low-energy B/C data require that 
the diffusion coefficient decreases with increasing energy; however, it is also possible
to achieve agreement with these data in \scenL\ for constant diffusion coefficient.

(5) Antiproton flux and \pbar/$p$ ratio seem to disfavor the low-energy
local source 
hypothesis (\scenL). Measurements of \pbar\ and/or \pbar/$p$ above 1~TeV may 
help to differentiate between the other scenarios.

(6) Radial component of the diffusive anisotropy of CR flux is too high in all 
scenarios, but the discrepancy is larger in \scenL, while \scenP\ predicts 
the lowest anisotropy. 
The PD model predicts a higher anisotropy than the D-R model.
Local sources may significantly affect the CR anisotropy, and therefore our 
simple analysis applies only to energy range unaffected by local sources.

(7) Data on the positron flux and positron fraction are inconsistent with
any of the scenarios, if all observed CR positrons are secondary.
However, if some of the detected positrons are produced in sources,
then only \scenH\ (local high-energy source) can account for
the observed positron excess at high energies.

(8) Finally, the \gray\ data are in agreement, within the uncertainty range, 
with all scenarios, including \scenR, even though the reference scenario 
does not agree with the new measurements for the CR proton and He spectra. 
\scenL\ slightly underpredicts the \gray\ flux below a few GeV.

Most specific physical models explaining the $p$/He ratio, spectral break and 
the dip fall into one of the scenarios studied in this paper, or their 
combination. 
Data from experiments such as the \fermilat\ and AMS-2 can be used to 
distinguish between some of these scenarios.

\acknowledgements
A.~V. and I.~V.~M. acknowledge support from NASA~grant~no.~NNX09AC15G.
T.~A.~P. acknowledges support from NASA~grant~no.~NNX10AE78G.
The authors are grateful to P.~Blasi, E.~Orlando, and A.~Strong for fruitful discussions and
to P.~Picozza and M.~Boezio for sharing and discussing preliminary PAMELA results.
The authors thank the anonymous referee for constructive comments.


\begin{deluxetable}{l p{5cm} *{6}{r} }
\tabletypesize{\scriptsize}
\tablecaption{Summary of Model Parameters, Diffusive-reacceleration (D-R) Model\label{table1}}
\tablehead{
& & \multicolumn{6}{c}{\it Calculations}\\ 
\cline{3-8}
\colhead{Parameter} & \colhead{Description} & \colhead{\it R} & \multicolumn{1}{r}{\it S$_1$ (S$_2$)} & \colhead{\it P} & \colhead{\it I} & \colhead{\it L} & \colhead{\it H} 
}
\startdata
{\sc Nucleon Injection}\smallskip\\
\quad $g_0$ (protons) &
         Injection index for $\rigidity<\Rinj$ & 
                  --1.90 & --1.90 & --1.90 & --1.90 & --1.90 & --1.90 \\
\qquad$\Rinj$ & 
         First break in CR injection spectrum, GV & 
                    11~~ & 11~~ & 11~~ & 11~~ & 11~~ & 11~~ \\
\quad $g_1$ (protons) & 
         Injection index for $\Rinj<\rigidity<\Rinjbr$ & 
                    --2.40 & --2.40 & --2.50 & --2.50 & --2.35 & --2.50 \\
\qquad$\Rinjbr$ & 
         Second break in CR injection spectrum, GV & 
                    \nodata & \nodata & \nodata & 300~ & \nodata & \nodata \\
\quad $g_2$ (protons) & 
         Injection index for $\Rinjbr<\rigidity$ & 
                    \nodata & \nodata & \nodata & --2.35 & \nodata & \nodata \medskip\\ 

\quad $\DeltapHe$ & 
         For nuclei, $g_{i}$($A>1$)=$g_{i}$(protons)$+\DeltapHe$ &
                   \nodata & \nodata & 0.07 & 0.07 & 0.07 & 0.07 \\ 
\quad $N_p$ & 
         Flux of protons at $\rigidity=10^3$~GV, in units 10$^{-12}$ cm$^{-2}$ sr$^{-1}$ s$^{-1}$ MeV$^{-1}$ &      
                   10.7 & 10.7 & 8.56 & 8.56 & 8.56 & 7.26 \\
\quad $n(\mathrm{H})/n_0(\mathrm{H})$ & 
Multiplication factor for the gas number density relative to standard gas maps &
                \nodata & ${1.0\quad\atop\quad(2.0)}$ & \nodata & \nodata & \nodata & \nodata \\
\quad $^4$He/$^1$H & Abundance of $^4$He relative to $^1$H in CR sources at 10$^3$~GeV~nucleon$^{-1}$. Abundances of other isotopes are proportional to $^4$He.&
                0.0686 &
                ${0.0910\quad \atop \quad (0.113)}$ &
                0.0842 &
                0.0932 &
                0.0944 &
                0.0830
\medskip \\
{\sc Electron Injection}\smallskip\\
\quad $\gez$ &
         Electron injection index for $\rigidity<\Rez$ & 
                  --1.60 & --1.60 & --1.60 & --1.60 & \nodata & --1.60 \\
\qquad$\Rez$ & 
         Low-energy break for electrons, GV & 
                    4~~ & 4~~ & 4~~ & 4~~ & \nodata & 4~~ \\
\quad $\geo$ &
         Injection index for $\Rez<\rigidity<\Reo$ & 
                  --2.50 & --2.50 & --2.70 & --2.70 & \nodata & --2.70 \\
\qquad$\Reo$ & 
         Intermediate energy break for electrons, GV & 
                    \nodata & \nodata & \nodata & 70 & \nodata & \nodata \\
\quad $\get$ & 
         Injection index for $\Reo<\rigidity<\Ret$ & 
                    \nodata & \nodata & \nodata & --2.33 & --2.33 & \nodata \\
\qquad$\Ret$ & 
         High-energy break for electrons, GV & 
                    $2\times 10^3$ & $2\times 10^3$ & $2\times 10^3$ & $2\times 10^3$ & $2\times 10^3$ & $2\times 10^3$ \\
\quad $\geh$ & 
         Injection index for $\Ret<\rigidity$ & 
                    5.0 & 5.0 & 5.0 & 5.0 & 5.0 & 5.0 \medskip\\ 
\quad $N_e$ & 
         Flux of protons at $\rigidity=R_e$, in units 10$^{-9}$ cm$^{-2}$ sr$^{-1}$ s$^{-1}$ MeV$^{-1}$ &      
                   1.100 & 1.100 & 1.100 & 1.166 & 5.15e-4 & 1.166 \\
\quad $R_e$ & 
         Rigidity for proton flux normalization, GV &      
                   24.8 & 24.8 & 24.8 & 24.8 & 300 & 24.8 \\

\medskip \\
{\sc Propagation}\smallskip\\
\quad $\valf$ & 
         Alfv\'{e}n speed &
                    32 & 25 & 32 & 32 & 32 & 32 \\
\quad $D_0$ & 
         Diffusion coefficient in 10$^{28}$ cm$^2$ s$^{-1}$ at $\rigidity$=4~GV  & 
                    5.75 & 4.00 & 5.75 & 5.75 & 5.75 & 5.75 \medskip\\
\quad $\delta_0$ & 
         $\delta$ in Equation (\ref{Dxx}) for $\rigidity<\Rdiflow$ & 
                    \nodata & \nodata & \nodata & \nodata & \nodata & \nodata \\
\qquad$\Rdiflow$ & 
         Low energy diffusion coefficient break, GV & 
                   \nodata & \nodata & \nodata & \nodata & \nodata & \nodata \\
\quad $\delta_1$ & 
         $\delta$ in Equation (\ref{Dxx}) for $\rigidity<\Rdiflow$ & 
                    0.30 & 0.30 & 0.30 & 0.30 & 0.30 & 0.30 \\
\qquad$\Rdifhigh$ & 
         High energy diffusion coefficient break, GV & 
                   \nodata & \nodata & 300.0 & \nodata & \nodata & \nodata \\
\quad $\delta_2$ & 
         $\delta$ for $\rigidity>\Rdifhigh$ & 
                   \nodata & \nodata & 0.15 & \nodata & \nodata & \nodata \\
\enddata 
\end{deluxetable}

\begin{deluxetable}{l p{5cm} *{6}{r} }
\tabletypesize{\scriptsize}
\tablecaption{Summary of Model Parameters, Plain Diffusion (PD) Model\label{table1PD}}
\tablehead{
& & \multicolumn{6}{c}{\it Calculations}\\ 
\cline{3-8}
\colhead{Parameter} & \colhead{Description} & \colhead{\it R} & \multicolumn{1}{r}{\it S$_1$ (S$_2$)} & \colhead{\it P} & \colhead{\it I} & \colhead{\it L} & \colhead{\it H} 
}
\startdata
{\sc Nucleon Injection}\smallskip\\
\quad $g_0$ (protons) &
         Injection index for $\rigidity<\Rinj$ & 
                  \nodata & \nodata & \nodata & \nodata & --2.40 & \nodata \\
\qquad$\Rinj$ & 
         First break in CR injection spectrum, GV & 
                    \nodata & \nodata & \nodata & \nodata & 54 & \nodata \\
\quad $g_1$ (protons) & 
         Injection index for $\Rinj<\rigidity<\Rinjbr$ & 
                  --2.10 & --2.10 & --2.26 & --2.26 & --2.05 & --2.26 \\
\qquad$\Rinjbr$ & 
         Second break in CR injection spectrum, GV & 
                    \nodata & \nodata & \nodata & 300 &  \nodata & \nodata \\
\quad $g_2$ (protons) & 
         Injection index for $\Rinjbr<\rigidity$ & 
                    \nodata & \nodata & \nodata & --2.05 & \nodata & \nodata \medskip\\ 

\quad $\DeltapHe$ & 
         For nuclei, $g_{i}$($A>1$)=$g_{i}$(protons)$+\DeltapHe$ &
                   \nodata & \nodata & 0.07 & 0.07 & 0.07 & 0.07 \\ 
\quad $N_p$ & 
         Flux of protons at $\rigidity=10^3$~GV, in units 10$^{-12}$ cm$^{-2}$ sr$^{-1}$ s$^{-1}$ MeV$^{-1}$ &      
                   10.7 & 10.7 & 8.56 & 8.56 & 8.56 & 6.93 \\
\quad $n(\mathrm{H})/n_0(\mathrm{H})$ & 
Multiplication factor for the gas number density relative to standard gas maps &
                \nodata & ${1.0\quad\atop\quad(2.0)}$ & \nodata & \nodata & \nodata & \nodata \\
\quad $^4$He/$^1$H & Abundance of $^4$He relative to $^1$H in CR sources at 10$^3$~GeV~nucleon$^{-1}$. Abundances of other isotopes are proportional to $^4$He.&
                0.0888 &
                ${0.111\quad \atop \quad (0.135)}$ &
                0.103 &
                0.115 &
                0.118 &
                0.0968
\medskip \\
{\sc Electron Injection}\smallskip\\
\quad $\gez$ &
         Electron injection index for $\rigidity<\Rez$ & 
                  --1.60 & --2.00 & --2.00 & 2.00 & \nodata & --2.00 \\
\qquad$\Rez$ & 
         Low-energy break for electrons, GV & 
                    4~~ & 4~~ & 4~~ & 4~~ & \nodata & 4~~ \\
\quad $\geo$ &
         Injection index for $\Rez<\rigidity<\Reo$ & 
                  --2.35 & --2.35 & --2.67 & --2.67 & \nodata & --2.67 \\
\qquad$\Reo$ & 
         Intermediate energy break for electrons, GV & 
                    \nodata & \nodata & \nodata & 61 & \nodata & \nodata \\
\quad $\get$ & 
         Injection index for $\Reo<\rigidity<\Ret$ & 
                    \nodata & \nodata & \nodata & --2.24 & --2.24 & \nodata \\
\qquad$\Ret$ & 
         High-energy break for electrons, GV & 
                    $2\times 10^3$ & $2\times 10^3$ & $2\times 10^3$ & $2\times 10^3$ & $2\times 10^3$ & $2\times 10^3$ \\
\quad $\geh$ & 
         Injection index for $\Ret<\rigidity$ & 
                    5.0 & 5.0 & 5.0 & 5.0 & 5.0 & 5.0 \medskip\\ 
\quad $N_e$ & 
         Flux of protons at $\rigidity=R_e$, in units 10$^{-9}$ cm$^{-2}$ sr$^{-1}$ s$^{-1}$ MeV$^{-1}$ &      
                   1.100 & 1.100 & 1.100 & 1.166 & 5.15e-4 & 1.166 \\
\quad $R_e$ & 
         Rigidity for proton flux normalization, GV &      
                   24.8 & 24.8 & 24.8 & 24.8 & 300 & 24.8 \\

\medskip \\
{\sc Propagation}\smallskip\\
\quad $\valf$ & 
         Alfv\'{e}n speed &
                    \nodata & \nodata & \nodata & \nodata & \nodata & \nodata \\
\quad $D_0$ & 
         Diffusion coefficient in 10$^{28}$ cm$^2$ s$^{-1}$ at $\rigidity$=4~GV  & 
                    3.00 & 2.20 & 3.00 & 3.00 & 1.30 & 3.00 \medskip\\
\quad $\delta_0$ & 
         $\delta$ in Equation (\ref{Dxx}) for $\rigidity<\Rdiflow$ & 
                    0.0 & 0.0 & 0.0 & 0.0 & 0.0 & 0.0 \\
\qquad$\Rdiflow$ & 
         Low energy diffusion coefficient break, GV & 
                   4.0 & 4.0 & 4.0 & 4.0 & 4.0 & 4.0 \\
\quad $\delta_1$ & 
         $\delta$ in Equation (\ref{Dxx}) for $\rigidity<\Rdiflow$ & 
                    0.60 & 0.60 & 0.60 & 0.60 & 0.60 & 0.60 \\
\qquad$\Rdifhigh$ & 
         High energy diffusion coefficient break, GV & 
                   \nodata & \nodata & 300.0 & \nodata & \nodata & \nodata \\
\quad $\delta_2$ & 
         $\delta$ for $\rigidity>\Rdifhigh$ & 
                   \nodata & \nodata & 0.37 & \nodata & \nodata & \nodata \\
\enddata 
\end{deluxetable}

\newpage

\newpage

\newpage

\newpage

\newpage

\end{document}